\documentclass{article}

\usepackage{arxiv}

\usepackage[utf8]{inputenc} 
\usepackage[T1]{fontenc}   
\usepackage[colorlinks=true, urlcolor=blue, linkcolor=blue, citecolor=blue,pdfborder={0 0 0}]{hyperref}      
\usepackage{url}           
\usepackage{booktabs}       
\usepackage{amsfonts}     
\usepackage{nicefrac}     
\usepackage{microtype}     
\usepackage{lipsum}	
\usepackage{graphicx}
\usepackage{array}
\usepackage{eqnarray,amsmath}
\usepackage{wrapfig}
\usepackage{mdwlist}
\usepackage{caption}
\usepackage{makecell}

\title{Transformer-CNN: Fast and Reliable tool for QSAR}

\author{
  Pavel Karpov\thanks{Helmholtz Zentrum M{\"u}nchen -- Research Center for Environmental Health (GmbH), Institute of Structural Biology, Ingolst{\"a}dter Landstra{\ss}e 1, D-85764 Neuherberg, Germany, \url{https://www.helmholtz-muenchen.de/stb/index.html}}\\
  Institute of Structural Biology\\
  Helmholtz Zentrum M{\"u}nchen,\\
  and BigChem GmbH, \\
  Germany, Munich \\
  \texttt{carpovpv@gmail.com} \\
   \And 
  Guillaume Godin\\
  Firmenich International SA,\\
  Research\&Development Division, \\
  Switzerland, Geneva \\
  \texttt{guillaume.godin@firmenich.com}
  \And
  Igor V.~Tetko \\
  Institute of Structural Biology\\
  Helmholtz Zentrum M{\"u}nchen,\\
  and BigChem GmbH, \\ 
  Germany, Munich \\ 
  \texttt{itetko@bigchem.de} \\
}

\begin{document}
\maketitle

\begin{abstract}

We present SMILES-embeddings derived from the internal encoder state of a Transformer~\cite{AttentionArticle} model trained to canonize SMILES as a Seq2Seq problem. Using a CharNN \cite{CharNN} architecture upon the embeddings results in higher quality interpretable QSAR/QSPR models on diverse benchmark datasets including regression and classification tasks. The proposed Transformer-CNN method uses SMILES augmentation for training and inference, and thus the prognosis is based on an internal consensus. That both the augmentation and transfer learning are based on embeddings allows the method to provide good results for small datasets. We discuss the reasons for such effectiveness and draft future directions for the development of the method. The source code and the embeddings needed to train a QSAR model are available on \url{https://github.com/bigchem/transformer-cnn}. The repository also has a standalone  program for QSAR prognosis which calculates individual atoms contributions, thus interpreting the model’s result. OCHEM~\cite{OCHEM} environment (\url{https://ochem.eu}) hosts the on-line implementation of the method proposed.

\end{abstract}


\section{Introduction}

Quantitative Structure-Activity (Property) Relationship (QSAR/QSPR) approaches find a nonlinear function, often modeled as an artificial neural network (ANN), that estimates the activity/property based on a chemical structure. In the past, most QSAR works heavily relied on descriptors~\cite{Todeschini} that represent in a numerical way some features of a complex graph structure of a compound. Amongst numerous families of descriptors, the fragment descriptors that count occurrences of a subgraph in a molecule graph, hold a distinctive status due to simplicity in the calculation. Moreover, there is a theoretical proof that one can successfully build any QSAR model with them~\cite{FragmentDescriptors}. Even a small database of compounds contains thousands of fragmental descriptors and some feature selection algorithm has traditionally been used to find a proper subset of descriptors for better quality, and to speed up the whole modeling process. Thus, feature selection in conjunction with a suitable machine learning method was key to success~\cite{FeatureSelection}. The rise of deep learning~\cite{RiseOfDeepLearning} allows us to bypass tiresome expert and domain-wise feature construction by delegating this task to a neural network that can extract the most valuable traits of the raw input data required for modeling the problem at hand \cite{NeurlaFingerprint,Coley}. 

\begin{figure}[t]
    \centering
    \includegraphics{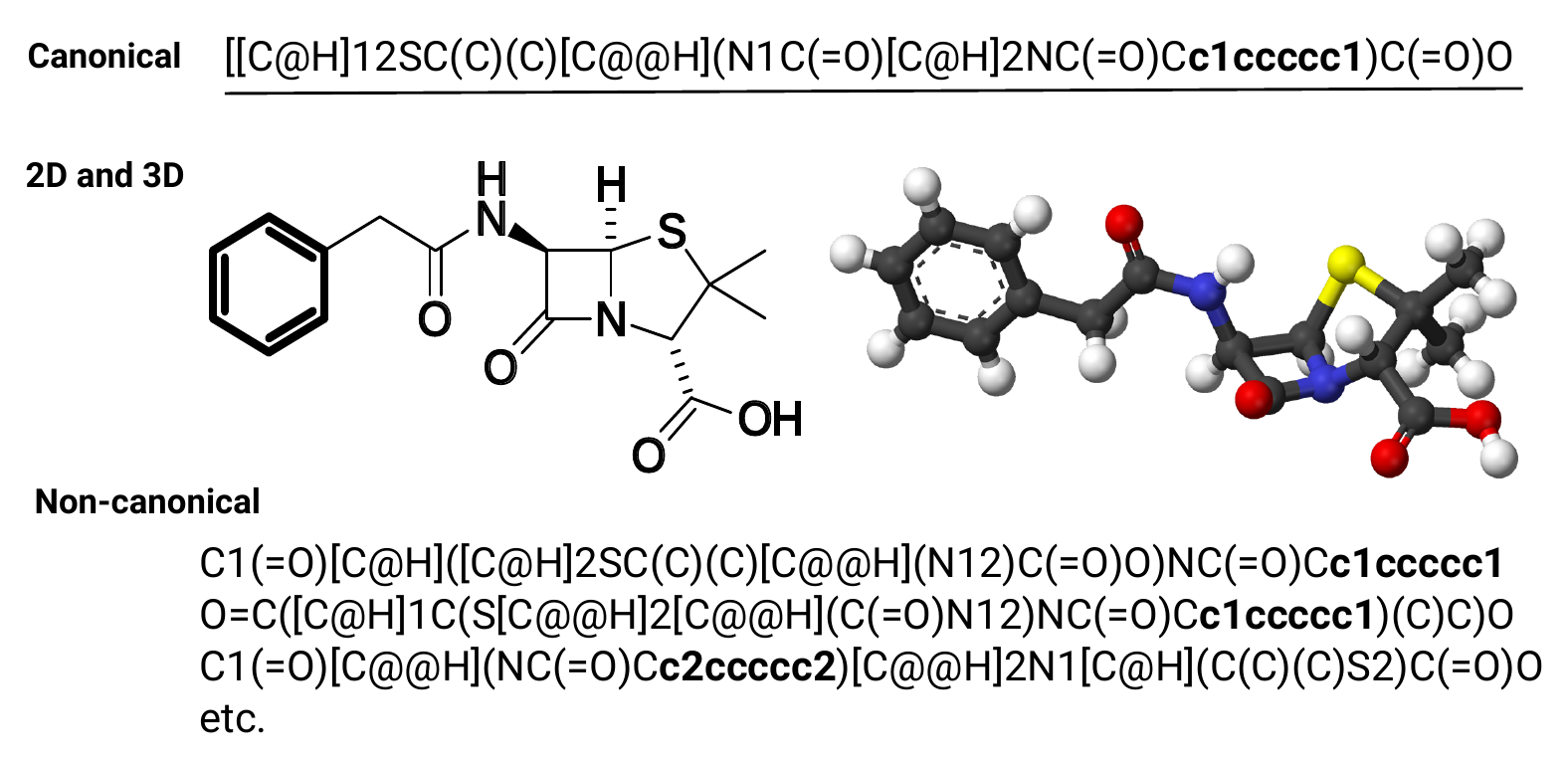}
    \caption{Benzylpenicillin canonical SMILES at the top, 2D and 3D structures derived from SMILES with OpenBabel~\cite{OpenBabel} in the middle, and three non-canonical SMILES examples at the bottom. A substructure of the phenyl ring is written in bold font.}
    \label{fig:structure}
    \vspace{-0.5cm}
\end{figure}

In this setting, the whole molecule as a SMILES-strings~\cite{Bombarelli,Augmentation} (Simplified Molecular Input Line Entry System)  or a graph~\cite{MessagePassing,EdgeAttention} serves as the input to the neural network. SMILES notation allows for the writing of any complex formula of an organic compound in a string facilitating storage and retrieval information about molecules in databases~\cite{Weininger}. It contains all information about the compound sufficient to derive the entire configuration (3D-structure) and has a direct connection to the nature of fragmental descriptors, Fig. \ref{fig:structure}, thus, making SMILES one of the best representation for QSAR studies. 

One of the first works exploiting direct SMILES input as descriptors used fragmentation of strings into groups of overlapping substrings forming a SMILES-like set or a hologram of a molecule~\cite{LINGO}. Within this approach, there was no need to derive a 2D/3D configuration of the molecule with subsequent calculation of descriptors keeping the quality of the models at the same level as with classical descriptors or even better.

SMILES strings are sequences of characters; therefore, they can be analyzed by machine-learning methods suitable for text processing, namely with convolutional and recurrent neural networks. After the demonstration of text understanding from character-level inputs~\cite{LeCunCharNN}, this technique was adopted in chemoinformatics~\cite{Augmentation,Smiles2Vec,LearningSmiles,chemNet,SmilesAttention}. Recently, we showed that the augmentation of SMILES (using canonical as well as non-canonical SMILES during model training and inference) increases the performance of convolutional models for regression and classification tasks~\cite{TetkoAugmentation}.

\begin{figure}[t]
    \centering
    \includegraphics[width=\textwidth]{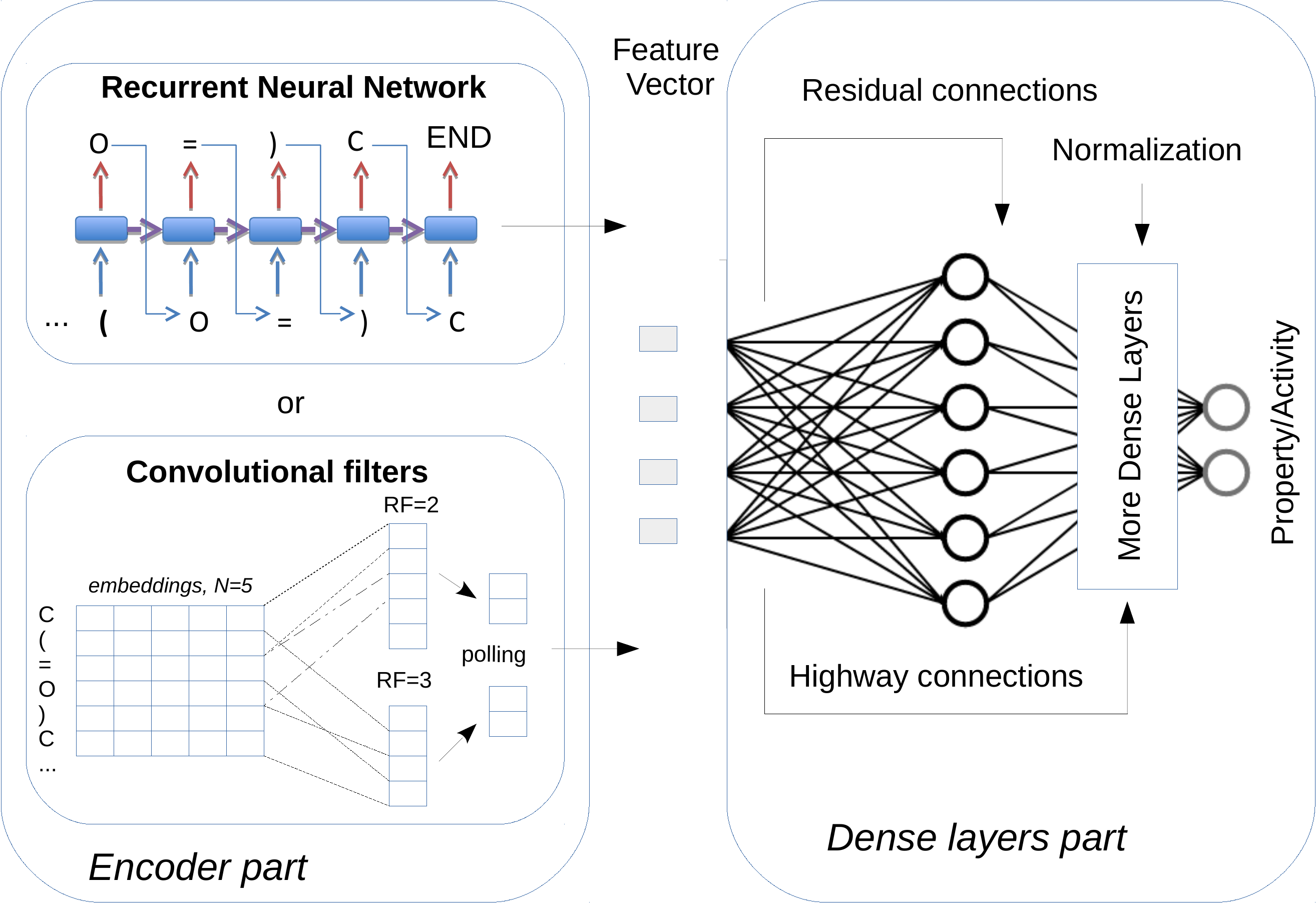}
    \caption{Scheme of modern QSAR models based on ANN. The encoder part (left) extracts main features of the input data by means of RNN (top) or convolutional layers (bottom). Then the feature vector as usual descriptors feeds to the dense layer part consisting of residual and highway connections, normalization layers, and dropouts.}
    \label{fig:encoder}
    \vspace{-5mm}
\end{figure}

Technically modern machine-learning models consist of two parts working together. The first part encodes the input data and extracts the most robust features by applying convolutional filters with different receptive fields (RF) or recurrent layers, whereas the second part directly builds the regular model based on these features using standard dense layers as building blocks (so called classical “MLP”), Fig.~\ref{fig:encoder}. Though powerful convolutional layers can effectively encode the input within its internal representation, usually one needs a considerable training dataset and computational resources to train the encoder part of a network.

The concept of embeddings mitigates the problem by using the pre-trained weights designed for image~\cite{ImageEmbeddings} or text processing~\cite{TextEmbeddings} tasks. It allows transfer learning from previous data and speeds up the training process for building models with significantly smaller datasets inaccessible for training from scratch. Typically, QSAR datasets contain only several hundreds of molecules, and SMILES-embeddings could improve models by developing better features.

One way of separately obtaining SMILES embeddings is to use classical autoencoder~\cite{Autoencoders} approach where the input is the same as the output. In the case of SMILES, however, it would be more desirable to explore a variety of SMILES belonging to the same molecule due to redundant SMILES grammar, Fig.~\ref{fig:structure}. We hypothesized that it is possible to train a neural network to conduct a SMILES canonicalization task in a Sequence-to-Sequence (Seq2Seq) manner like a machine translation problem, where on the left side are non-canonical SMILES, and on the right side are their canonical equivalents. Recently, Seq2Seq was successfully applied to translation from InChi~\cite{InChi} codes to SMILES (Inchi2Sml) as well as from SMILES arbitrary to canonical SMILES (Sml2canSml), and to build QSAR models on extracted latent variables~\cite{smi2Inchi}. 

The state-of-the-art neural architecture for machine translation consists of stacked Long Short-Term Memory (LSTM) cells~\cite{LSTM}. The training process for such networks inherently has all kinds of Recurrent Neural Networks difficulties, e.g., vanishing gradients, and the impossibility of parallelization. Recently, a Transformer model~\cite{AttentionArticle} was proposed where all recurrent units are replaced with convolutional and element-wise feed-forward layers. The whole architecture shows a significant speed-up during training and inference with improved accuracy over translation benchmarks. The Transformer model was applied for prediction of reaction outcomes~\cite{Schwaller} and for retrosynthesis~\cite{KarpovRetro}.

Modern machine learning architectures although demonstrating incredible performance still lack interpretability. Explaining the reasons for a particular prediction of a model avoids "Clever Hans" predictors with spurious or non-relevant correlations~\cite{Samek2019} and foster trust and verifiability. One of the promising methods to open a "black box" uses the Layer-wise Relevance  Propagation (LRP) algorithm~\cite{Montavon}, which splits the overall predicted value to a sum of contributions of individual neurons. In this method, the sum of relevance of all neurons of a layer, including the bias neuron, is kept constant. Propagation of the relevance from the last layer to the input layer allows the evaluation of the contributions of particular input features in to select the most relevant features for the whole training set~\cite{TetkoPruning} or to explain the individual neural network prediction~\cite{Montavon}. We apply the LRP method for an explanation of individual results, checking the model get results for the right reason.

Our contributions in the article are as follows:
\begin{itemize*}
\item presenting a concept of dynamic SMILES embeddings that may be useful for a wide range of cheminformatics tasks;
\item scrutinizing CharNN models based on these embeddings for regression and classification tasks and show that the method outperforms the state-of-the-art models;
\item Interpretation of the model based on LRP method;
\item our implementation as well as source codes and SMILES-embeddings are available on  \url{https://github.com/bigchem/transformer-cnn}. We also provide ready-to-use implementation on \url{https://ochem.eu} within  the OCHEM~\cite{OCHEM} environment and a standalone program for calculating properties and explaining the results.

\end{itemize*}

\section{Methods}
\subsection{SMILES canonicalization model}
\subsubsection{Dataset}

\begin{figure}[h!]
    \centering
    \includegraphics[width=\textwidth]{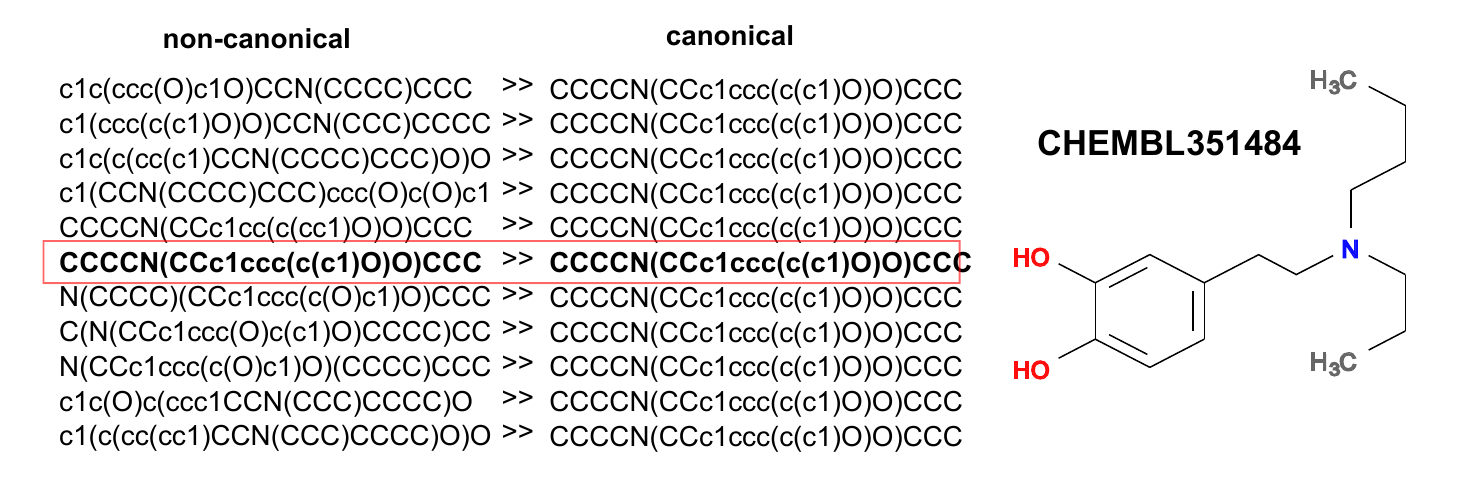}
    \caption{Example of the data in the training file for canonicalization model of a small molecule CHEMBL351484. Every line contains a pair of non-canonical (left) and canonical (right) separated by “>>”. One line has identical SMILES on both sides, stressed with the red box.}
    \label{fig:noncanon}
\end{figure}

To train the ANN to perform SMILES canonicalization, we used the ChEMBL database~\cite{ChEMBL} with SMILES strings of length less than or equal 110 characters (>93\% of the entire database). The original dataset was augmented 10 times up to 17,657,995 canonicalization pairs written in reactions format separated by ‘>>’. Each pair contained on the left side a non-canonical, and on the right side -- a canonical SMILES for the same molecule. Such an arrangement of the training dataset allowed us to re-use the previous Transformer code, which was originally applied for retrosynthetic tasks~\cite{KarpovRetro}. For completeness, we added for every compound a line where both left and right sides were identical, i.e. canonical SMILES,  Fig.~\ref{fig:noncanon}. Thus each molecule was present in the training set 11 times. If a molecule had tautomeric forms then all of them were accounted for as separate entries in the training data file.

\subsubsection{Model Input}

Seq2Seq models use one-hot encoding vector for the input. Its values are zero everywhere except the position of the current token which is set to one. Many works on SMILES use tokenization procedure~\cite{Segler,Gupta} that combines some characters, for example ‘B’ and ‘r’ to one token ‘Br’. Other rules for handling most common two-letters elements, charges, and stereochemistry also are used for preparing the input for the neural network. According to our experience, the use of more complicated schemes instead of simple character-level tokenization did not increase the accuracy of models~\cite{KarpovRetro}. Therefore a simple character-level tokenization was used in this study. The vocabulary of our model consisted of all possible characters from ChEMBL dataset and has 66 symbols:

\begin{quote}
\texttt{" \textasciicircum\#\%()+-.\/0123456789=@ABCDEFGHIKLMNOPRSTVXYZ[\textbackslash\textbackslash]abcdefgilmnoprstuy\$"}
\end{quote}

Thus, the model could handle the entire diversity of drug-like compounds including stereochemistry, different charges, and inorganic ions. Two special characters were added to the vocabulary: "\textasciicircum" to indicate the start of the sequence, and "\$" to inform the model of the end of data input. 

\subsubsection{Transformer model}

The canonicalization model used in this work was based upon a Transformer architecture consisting of two separate stacks of layers for the encoder and the decoder, respectively. Each layer incorporated some portion of knowledge written in its internal memory (V) with indexed access by keys (K). When new data arrived (Q), the layer calculated attention and modified the input accordingly (see the original work on Transformers~\cite{AttentionArticle}), thus, forming the output of the self-attention layer and weighting those parts that carry the essential information. Besides a self-attention mechanism, the layer also contained several position-wise dense layers, a normalization layer, and residual connections~\cite{AttentionArticle,AnnotatedTransformer}. Our model utilized a three layer architecture of Transformer with 10 blocks of self-attention, i.e. the same one as used in our previous study~\cite{KarpovRetro}. After the encoding process was finished, the output of the top encoder layer contained a representation of a molecule suitable for decoding into canonical SMILES. In this study we used this representation as a well-prepared latent representation for QSAR modeling. 

Tensorflow v1.12.02~\cite{tensorflow} was used as machine-learning framework to develop all parts of the Transformer, whereas RDKit v.2018.09.2~\cite{rdkit} was used for SMILES canonicalization and augmentation. 

\subsection{QSAR model}

\begin{figure}[h]
    \centering
    \includegraphics[width=\textwidth]{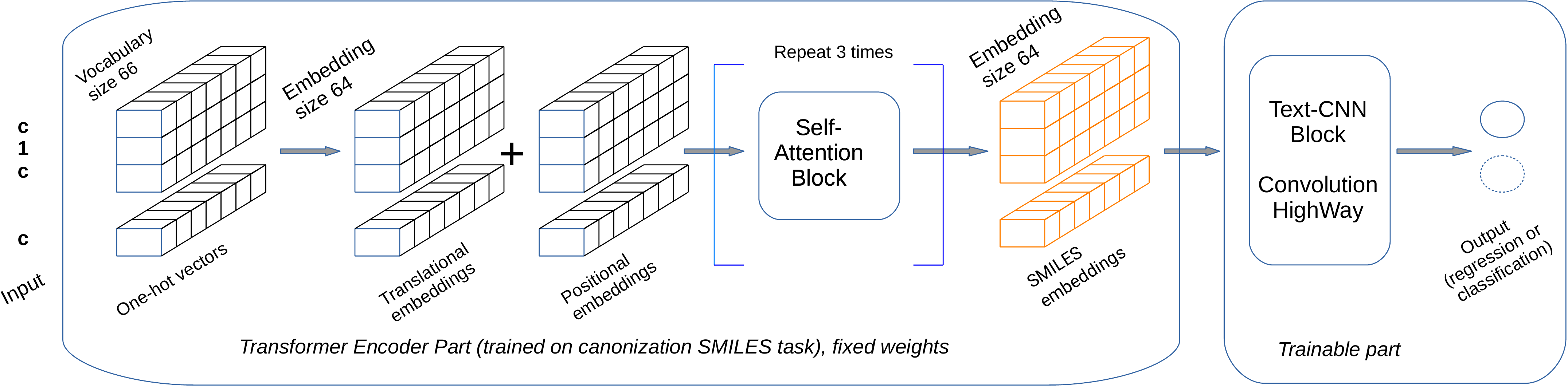}
    \caption{The architecture of the Transformer-CNN network.}
    \label{fig:model}
\end{figure}

We call the output of the Transformer's encoder part a dynamic SMILES-embedding, Fig.~\ref{fig:model}. For a molecule with N-characters, the encoder produces the matrix with dimensions (N, EMBEDDINGS). Though technically this matrix is not an embedding because equivalent characters have different values depending on position and surroundings, it can be considered so due to its role: to convert an input one-hot raw vectors to real-value vectors in some latent space. Because these embeddings have variable lengths, we used a series of 1D convolutional filters as implemented in DeepChem~\cite{deepchem} TextCNN method (\url{https://github.com/deepchem}). 

Each convolution had a kernel size from the list [1, 2, 3, 4, 5, 6, 7, 8, 9, 10, 15, 20] and produced the following number of filters [100, 200, 200, 200, 200, 100, 100, 100, 100, 100, 160, 160], respectively. After a GlobalMaxPool operation and the subsequent concatenation of the pooling results, the data went through Dropout~\cite{dropout} (rate=0.25), Dense(N=512), Highway~\cite{highway} layers, and, finally, converted to the output layer which consisted of only one neuron for regression and two neurons for classification tasks. The weights of the Transformer’s part were frozen in all experiments. All models used the Adam optimizer with Mean Squared Error or Binary Cross-Entropy loss depending on the problem at hand.  A fixed learning rate $\lambda = 10^{-4}$ was used. Early-stopping was used to prevent overfitting, to select a best model, and to reduce training time. OCHEM calculations were performed using canonical SMILES as well as ten-fold augmented SMILES during both training and prognosis. This number of SMILES augmentations was found to be an optimal one in our previous study~\cite{TetkoAugmentation}. An average value of the individual predictions for different representation of the same molecule was used as the final model prediction to calculate statistical parameters. 

The same five-fold cross-validation procedure was used to compare the models with the results of our previous study~\cite{TetkoAugmentation}. The coefficients of determination~\cite{Draper}:

\begin{equation}
    r^2 = 1 - {{SS_{res}}\over{SS_{tot}}},
\end{equation}

where $SS_{tot}$  is total variance of data and $SS_{res}$  is residual unexplained variance of data was used to compare regression models and Area Under the  Curve (AUC) was used for classification tasks.

\subsection{Validation datasets}

We used the same datasets (9 for regression and 9 for classification) that were exploited in our previous studies~\cite{TetkoAugmentation,Augmentation}. Short information about these sets as well as links to original works are provided in Table~\ref{tbl:datasets}. The datasets are available on the OCHEM environment on \url{https://ochem.eu}.

\begin{table}
\caption{Descriptions of datasets used in the work.}
\centering
\begin{tabular}{p{1cm}p{5cm}p{1cm}|p{1cm}p{5cm}p{1cm}}
\toprule
    Code & Description & Size & Code & Description & Size \\
    \midrule
    \multicolumn{3}{c}{Regression tasks} & \multicolumn{3}{|c}{Classification tasks} \\ \midrule 
    MP & Melting Point~\cite{MP} & \multicolumn{1}{r|}{19,104} & HIV & Inhibition of HIV replication~\cite{benchmark} & \multicolumn{1}{r}{41,127} \\ 
    BP & Boiling Point~\cite{BP} & \multicolumn{1}{r|}{11,893} & AMES & Mutagenicity~\cite{ames} & \multicolumn{1}{r}{6,542} \\
    BCF & Bioconcentration factor~\cite{BP} & \multicolumn{1}{r|}{378} & BACE & Human $\beta$-secretase 1 (BACE-1) inhibitors~\cite{benchmark} & \multicolumn{1}{r}{1,513} \\
    FreeSolv & Free solvation energy~\cite{benchmark} & \multicolumn{1}{r|}{642} & Clintox & Clinical trial toxicity~\cite{benchmark} & \multicolumn{1}{r}{1,478} \\
    LogS & Solubility~\cite{Tetkosolubility} & \multicolumn{1}{r|}{1,311} & Tox21 & In-vitro toxicity~\cite{benchmark} &  \multicolumn{1}{r}{7,831} \\
    Lipo & Lipophilicity~\cite{TetkoLipo} & \multicolumn{1}{r|}{4,200} & BBBP & Blood-brain barrier~\cite{benchmark} &  \multicolumn{1}{r}{2,039} \\
    BACE & IC50 of human $\beta$-secretase 1 (BACE-1) inhibitors~\cite{benchmark} & \multicolumn{1}{r|}{1,513} & JAK3 & Janus kinase 3 inhibitor~\cite{jak3} &  \multicolumn{1}{r}{886} \\
    DHFR & Dihydrofolate reductase inhibition~\cite{dhfr} & \multicolumn{1}{r|}{739} & BioDeg & Biodegradability~\cite{biodeg} &  \multicolumn{1}{r}{1,737} \\
    LEL & Lowest effect level~\cite{lel} & \multicolumn{1}{r|}{483} & RP AR & Endocrine disruptors~\cite{rpar} &  \multicolumn{1}{r}{930} \\
    \bottomrule
\end{tabular}
\label{tbl:datasets}
\vspace{-3mm}
\end{table}

\section{Results and discussion}
\subsection{SMILES canonicalization model}

The Transformer model was trained for 10 epochs with the learning rate changing according to the formula:
\begin{wrapfigure}{r}{0.46\textwidth}
    \vspace{-2mm}
    \centering
    \includegraphics[width=0.46\textwidth]{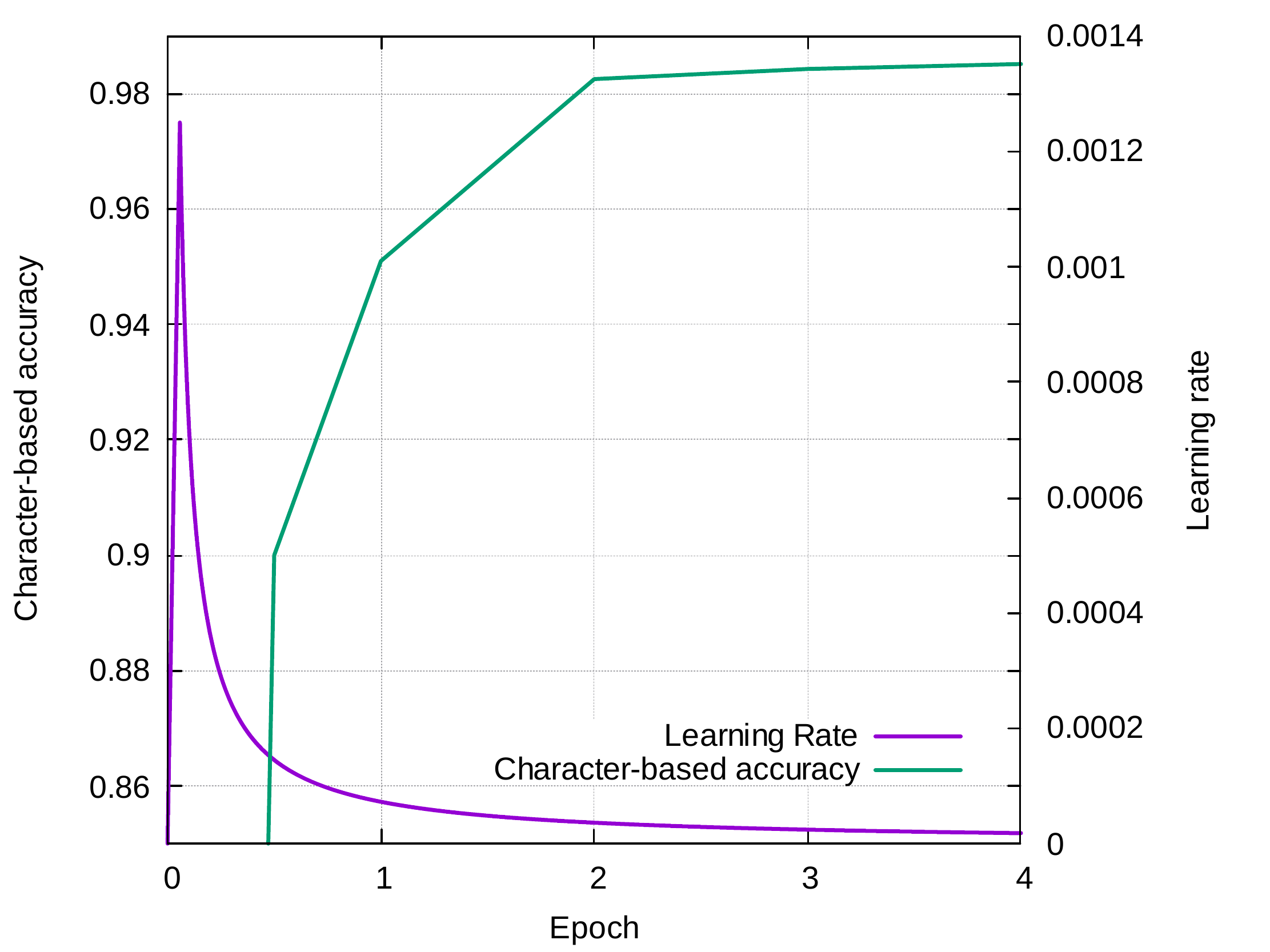}
    \captionof{figure}{Learning curves: 1) learning rate schedule (axes bottom and right), and 2) character-based accuracy (axes bottom and left) on the training dataset for the first four epochs.}
    \label{fig:curves}
    \vspace{-8mm}
\end{wrapfigure}
\begin{equation}
    \lambda = factor * { {min(1.0, { {step} \over{warmup} })} \over{ max(step, warmup) } },
\end{equation}

where $factor = 20$, $warmup = 16,000$ steps, and if $\lambda < 10^{-4}$ then $\lambda = 10^{-4}$. The settings for the learning rate were similar to those used in our retro-synthesis study. Each epoch contained 275,907 steps (batches). No early-stopping or weight-averaging was applied. Learning curves are shown in Fig. \ref{fig:curves}.

\begin{wrapfigure}{r}{0.46\textwidth}
\vspace{-5mm}
\captionof{table}{Validation of canonicalization model}
\centering
\begin{tabular}{p{1.5cm}p{1.3cm}p{3.3cm}} \toprule
Strings & All & Correctly canonicalized \\ \midrule
All & \multicolumn{1}{r}{500,000} & \multicolumn{1}{r}{418,233 (\textbf{83.6\%})} \\
Stereo & \multicolumn{1}{r}{77,472} & \multicolumn{1}{r}{28,821 (37,2\%)} \\
Cis/trans & \multicolumn{1}{r}{54,727} & \multicolumn{1}{r}{40,483 (73,9\%)} \\
\bottomrule 
\end{tabular}
\label{tbl:generator}
\end{wrapfigure}To validate the model, we sampled 500,000 ChEMBL-like SMILES (only 8,617 (1.7\%) of them were canonical) from a generator~\cite{xia} and checked how accurately the model can restore canonical SMILES for these molecules. We intentionally selected the generated SMILES keeping in mind possible applications of the proposed method in artificial intelligence-driven pipelines of de-novo drug development. The model correctly canonicalized 83,6\% of all samples, Table \ref{tbl:generator}.

\subsection{QSAR modeling}

\begin{figure}[t!]
    \centering
    \includegraphics[width=\textwidth]{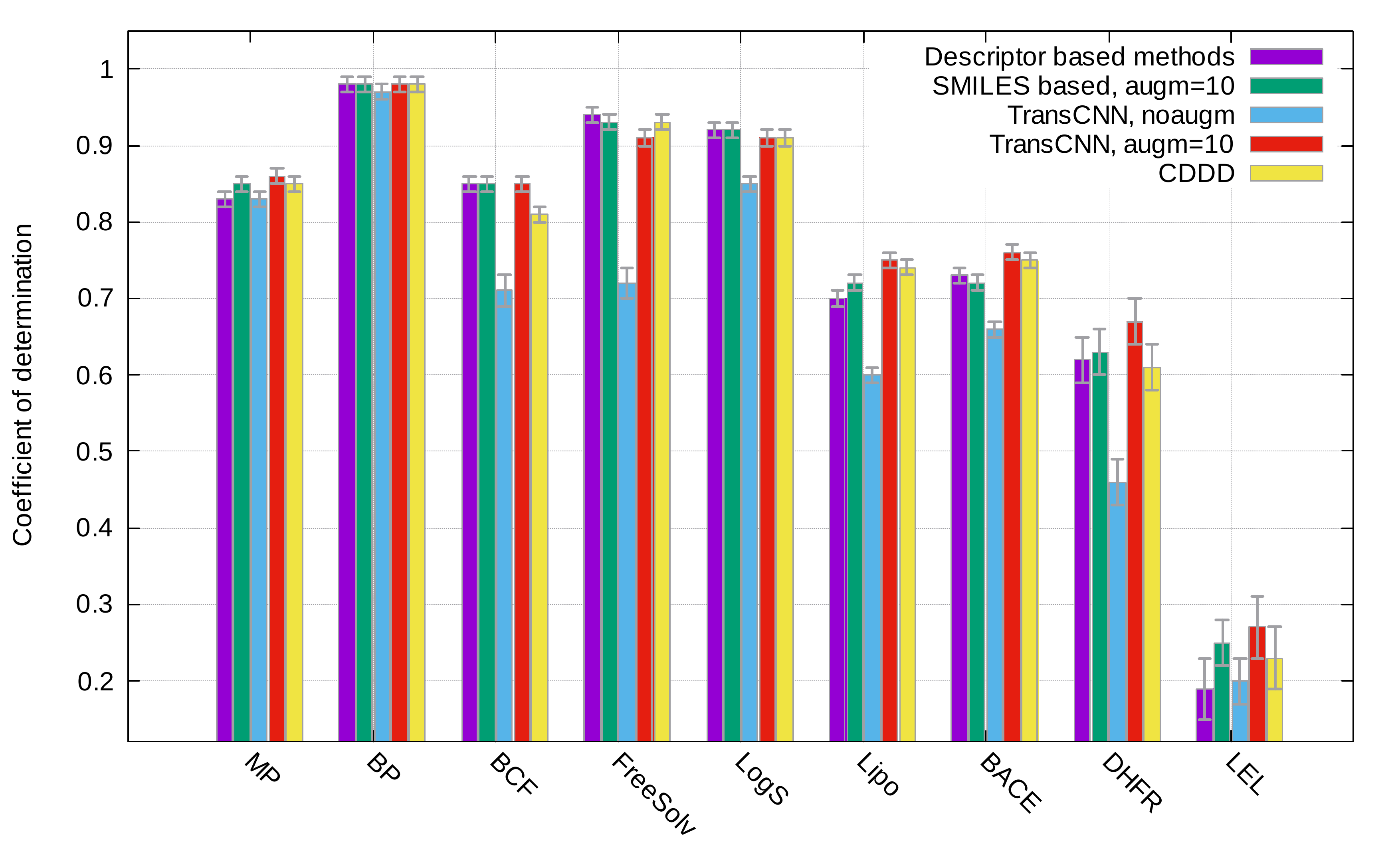}
    \caption{Coefficient of determination, $r^2$, calculated for regression sets (higher values are better).}
    \label{fig:regression}
\end{figure}

\begin{figure}[t!]
    \centering
    \includegraphics[width=\textwidth]{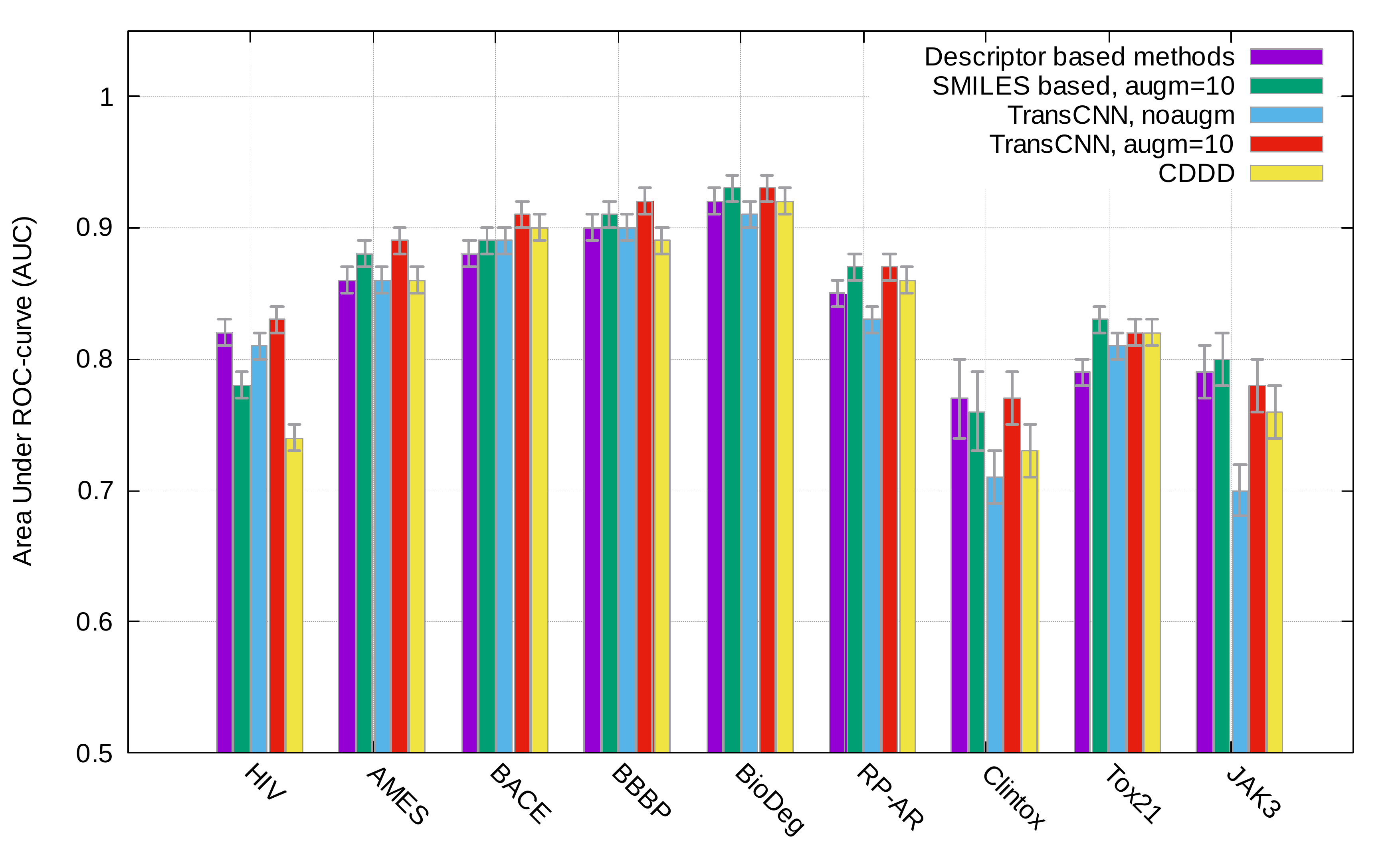}
    \caption{AUC calculated for classification sets (higher values are better).}
    \label{fig:classification}
\end{figure}

For the QSAR modeling the saved embedding was used. The training was done using a fixed learning rate $\lambda$ = 0.001 for n=100 epochs. Early stopping with 10\% randomly selected SMILES was used to identify the optimal model. Table \ref{tbl:regression}\footnote{We omitted the standard mean errors, which are 0.01 or less, for the reported values.}, Fig. \ref{fig:regression} compares results for regression datasets while Table \ref{tbl:classification}, Fig. \ref{fig:classification} compares classification tasks. The standard mean errors of the values were calculated using a bootstrap procedure as explained elsewhere~\cite{biodeg}.

With an exception of a few datasets, the proposed method provided similar or better results than those calculated using descriptor-based approaches as well as the other SMILES-based approaches investigated in our previous study~\cite{TetkoAugmentation}. The data augmentation was critically important for the Transformer-CNN method to achieve its high performance. We used augmentation n=10, i.e., 10 SMILES were randomly generated and used for model development and application, which was found optimal in the aforementioned previous study.

\begin{table}[t]
    \caption{Coefficient of determination, $r^2$, calculated for regression sets (higher values are better).}
    \centering
    \begin{tabular}{p{1.5cm}p{2.5cm}p{2.5cm}p{2.5cm}p{2.5cm}p{2.5cm}} \toprule
Dataset & Descriptor based methods & SMILES based (augm=10)~\cite{TetkoAugmentation} & Transformer-CNN, no augm. & Transformer-CNN, augm=10 & CDDD ~~~~~~~~~~~~~~~~ descriptors~\cite{smi2Inchi} \\ \midrule 
MP & 0.83 & 0.85 & 0.83 & 0.86 & 0.85 \\
BP & 0.98 & 0.98 & 0.97 & 0.98 & 0.98 \\
BCF & 0.85 & 0.85 & 0.71 $\pm$ 0.02 & 0.85 & 0.81 \\
FreeSolv & 0.94 & 0.93 & 0.72 $\pm$ 0.02 & 0.91 & 0.93 \\
LogS & 0.92 & 0.92 & 0.85 & 0.91 & 0.91 \\ 
Lipo & 0.7 & 0.72 & 0.6 & 0.73 & 0.74 \\
BACE & 0.73 & 0.72 & 0.66 & 0.76 & 0.75 \\
DHFR & 0.62 $\pm$ 0.03 & 0.63 $\pm$ 0.03 & 0.46 $\pm$ 0.03 & 0.67 $\pm$ 0.03 & 0.61 $\pm$ 0.03 \\
LEL & 0.19 $\pm$ 0.04 & 0.25 $\pm$ 0.03 & 0.2 $\pm$ 0.03 & 0.27 $\pm$ 0.04 & 0.23 $\pm$ 0.04 \\ 

\bottomrule
    \end{tabular}
    \label{tbl:regression}
\end{table}

Similar to Transformer-CNN the Sml2canSml used an internal representation, which was developed by mapping arbitrary SMILES to canonical SMILES. The difference was that Sml2canSml generated a fixed set of 512 latent variables (CDDD descriptors), while the Transformer-CNN representation had about the same length as the initial SMILES. Sml2canSml CDDD could be used as descriptors for any traditional machine learning methods while Transformer-CNN required convolutional neural networks to process the variable length output and to correlate it with the analysed properties. Sml2canSml was added as CDDD descriptors to OCHEM. These descriptors were analysed by the same methods as used in the previous work, i.e., LibSVM~\cite{libsvm}, Random Forest ~\cite{randomforest}, XGBoost~\cite{xgboost} as well as by Associative Neural Networks (ASNN)~\cite{asnn} and Deep Neural Networks~\cite{sosnin}. Exactly the same protocol, 5-fold cross-validation, was used for all calculations. The best performance using the CDDD descriptors was obtained by ASNN and LibSVM methods, which contributed models with the highest accuracy for seven and five datasets respectively (LibSVM method provided the best performance in the original study). Transformer-CNN provided better or similar results compared to the CDDD descriptors for all datasets with an exception of Lipo and FreeSolv. It should be also mentioned that CDDD descriptors could only process molecules which satisfy the following conditions:

\begin{itemize*}
\item logP $\in$ (-5,7),
\item mol\_weight $\in$ (12,600),
\item num\_heavy\_atoms $\in$ (3, 50),
\item molecule is organic.
\end{itemize*}

\begin{table}[t]
    \caption{AUC calculated for classification sets (higher values are better).}
    \centering
    \begin{tabular}{p{1.5cm}p{2.5cm}p{2.5cm}p{2.5cm}p{2.5cm}p{2.5cm}} \toprule
Dataset & Descriptor based methods & SMILES based (augm=10)~\cite{TetkoAugmentation} & Transformer-CNN, no augm. & Transformer-CNN, augm=10 & CDDD ~~~~~~~~~~~~~~~~ descriptors~\cite{smi2Inchi} \\ \midrule 
HIV & 0.82 & 0.78 & 0.81 & 0.83& 0.74 \\
AMES & 0.86 & 0.88 & 0.86 & 0.89 & 0.86 \\
BACE & 0.88 & 0.89 & 0.89 & 0.91 & 0.9 \\ 
Clintox & 0.77 $\pm$ 0.03 & 0.76 $\pm$ 0.03 & 0.71 $\pm$ 0.02 & 0.77 $\pm$ 0.02 & 0.73 $\pm$ 0.02 \\
Tox21 & 0.79 & 0.83 & 0.81 & 0.82 & 0.82 \\ 
BBBP & 0.90 & 0.91 & 0.9 & 0.92 & 0.89 \\
JAK3 & 0.79 $\pm$ 0.02 & 0.8 $\pm$ 0.02 & 0.70 $\pm$ 0.02 & 0.78 $\pm$ 0.02 & 0.76 $\pm$ 0.02 \\
BioDeg & 0.92 & 0.93 & 0.91 & 0.93 & 0.92 \\
RP AR & 0.85 & 0.87 & 0.83 & 0.87 & 0.86 \\
\bottomrule
    \end{tabular}
    \label{tbl:classification}
\end{table}

These limitations appeared due to the preparation of the training set to develop the Sml2canSml encoder. The limitations resulted in the exclusion of a number of molecules, which failed one or several of the above conditions. Contrary to the Sml2canSml encoder, we trained Transformer-CNN with very diverse molecules from ChEMBL and thus the developed models could be applied to any molecule which can be processed by RDKiT. The exclusion of molecules for which CDDD descriptors failed to be calculated did not significantly change the results of Transformer models: some models improved while others decreased their accuracy for \textasciitilde 0.01 respective performance values. For example, for Lipo and FreeSolv sets the accuracy of the Transformer-CNN model increased to $r^2$ = 0.92 and 0.75 respectively, while for BBB the AUC  decreased to 0.91.

\subsection{Interpretability of the model}

Layer-Wise Relevance propagation was used to interpret the models. For gated connections (in HighWay block) we implemented the signal-take-all redistribution rule~\cite{arras} while all other Dense and Convolutional layers were well fitted in the LRP framework~\cite{Montavon} without any adaptation. In this work, we stopped the relevance propagation on the output of the Transformer’s encoder which is position-wise. It should be noted that we froze the encoder part of the network during QSAR model training. Summing up all the individual features for each position in the SMILES string calculated its contribution to the final result. If the LRP indicated a reasonable explanation of the contributions of fragments then one can trust that the model made predictions based on detected fundamental structure-property relationships. For explanation we selected classification (AMES mutagenicity) and regression (water solubility) models.       

\subsubsection{AMES mutagenicity}

The AMES test is a widely used qualitative test to determine the mutagenic potential of a molecule, from which extensive structural alerts collections were derived~\cite{mutagen}. Examples of these alerts are aromatic nitros, N-oxides, aldehydes, monohaloalkenes, quinones, etc. A QSAR model for AMES had to pay special attention to these and similar groups to be interpretable and reliable. The Transformer-CNN model built on 6542 endpoints (3516 mutagenic and 3026 nonmutagenic) results in AUC = 0.89, Table~\ref{tbl:classification}. 

The structure of 1-Bromo-4-nitrobenzene gave the positive AMES test. The output of the LRP procedure for one of possible SMILES for this compound, namely 1c([N+]([O-])=O)ccc(c1)Br, is shown in Table \ref{tbl:lrp}\footnote{All zero values were all less than $10^{-5}$.}.

According to the LRP, the relevance was constant during the propagation:

\begin{equation}
y = R = f(x) = \sum_{l \in (L)} R_l = \sum_{l \in (L-1)} R_l = \sum_{l \in (L-2)} R_l = \dots = \sum_{l \in (1)} R_l.
\end{equation}

Here (L) stood for a set of neurons in the last, (L-1) -- in the layer before the last layer, and so on. Each layer in the Transformer-CNN network contained biases (B), and thus some relevance dissipated on them. Therefore the above equation was correced to:

\begin{equation}
\sum_{l \in (L)} R_l = \sum_{l \in (L-1)} R_l + B.
\end{equation}

\begin{table}[b]
    \caption{Local relevance conservation for c1c([N+]([O-])=O)ccc(Br)c1.}
    \centering
    \begin{tabular}{p{2.5cm}>{\centering\arraybackslash}p{3cm}>{\centering\arraybackslash}p{2.5cm}>{\centering\arraybackslash}p{3.2cm}>{\centering\arraybackslash}p{3.2cm}} \toprule 
Layer & Relevance, $R_{(L+1)}$ & Relevance, $R_{(L)}$ & $\Delta = R_{(L+1)} - R_{(L)}$ & $\Delta / R_{(L+1)} * 100$\% \\ \midrule 
Result & 0.9812 & - & - & - \\
HighWay Output & 0.9812 & 0.9300 & 0.0512 & 5.21 \\
HighWay Input & 0.9300 & 0.7227 & 0.2073 & 22.3 \\
DeMaxPool & 0.7227 & 0.7371 & -0.0144 & -1.98 \\
~Conv1 & 0.0090 & 0.0117 & -0.0027 & -30.1 \\
~Conv2 & 0.1627 & 0.1627 & 0 & 0 \\
~Conv3 & -0.0443 &  -0.0443 & 0 & 0 \\
~Conv4 & 0.0191 &  0.0191 &  0 &  0 \\
~Conv5 & -0.0984 & -0.0984 & 0 &  0 \\
~Conv6 & -0.0136 & -0.0136 & 0 &  0 \\
~Conv7 & 0.0806 & 0.0806 & 0 & 0 \\
~Conv8 & 0.0957 & 0.0957 & 0 & 0 \\
~Conv9 & 0.1528 & 0.1528 &  0 &  0 \\
~Conv10 &  0.0845 &  0.0845 & 0 &  0 \\
~Conv15 & 0.1038 & 0.1038 & 0 &  0 \\
~Conv20 &  0.1851 &  0.1851 &  0 & 0 \\ \midrule
\textbf{Total} & \textbf{0.98119} & \textbf{0.7398} & \textbf{0.2414} & \textbf{24.6} \\
    \bottomrule 
    \end{tabular}
    \label{tbl:lrp}
\end{table}

We calculated how much of the relevance was taken by biases and reported these values in the output of the ochem.py script. Table~\ref{tbl:lrp} clearly shows that 24.6\% of the output signal was taken by biases and 75.4\% were successfully propagated to position-wise layers, which we used to interpret the model. If less than 50\% of the signal came to the input, it may indicate an applicability domain problem or technical issues with relevance propagation. In these cases the interpretation could be questioned. 

\begin{figure}[t]
    \centering
    \includegraphics[width=\textwidth]{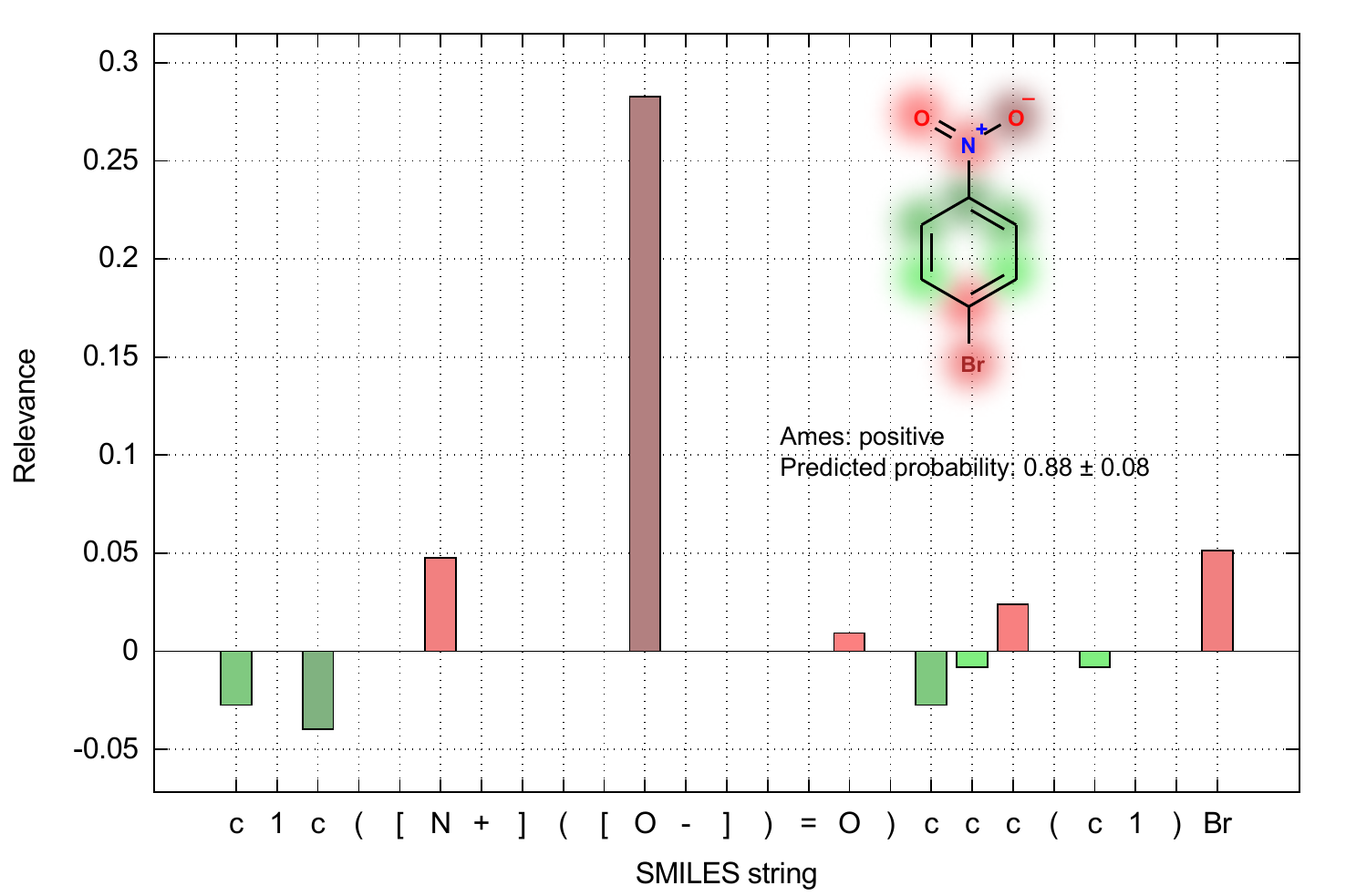}
    \caption{Visualization of atom contributions, in the case of a mutagenic compound. The red color stands for mutagenic alerts, color green against it.}
    \label{fig:ames}
\end{figure}

Iterating through all non-hydrogen atoms, the interpretation algorithm picked up an atom and drew a SMILES from it. Thus, every molecule had a corresponding set of SMILES equal to the number of atoms. The LRP was used for every SMILES, and then the individual predictions were averaged for the final output. 1-Bromo-4-nitrobenzene was predicted as mutagenic with the score 0.88. Impacts of the atoms on the property is depicted in Fig.~\ref{fig:ames}. The model predicted this compound as mutagenic because of the presence of nitro and halogen benzene moieties. Both are known to be structural alerts for mutagenicity~\cite{mutagen}. Charged oxygen provided a bigger impact than the double bonded one in the nitro group because its presence contributed to the mutagenicity for nitro and also for N-oxide compounds.

\subsubsection{Aqueous solubility}

Solubility is a crucial property in drug-development. To have a fast, robust, and explainable tool for its prediction and interpretation is highly desirable by both academia and industry. The Transformer-CNN model built on 1311 compounds had the following statistics: $q^2$ = 0.92 and RMSEp=0.57~\cite{xia}. For demonstration of its interpretability we choose haloperidol -- a well-known antipsychotic drug with 14 mg/l  water solubility~\cite{solub}. 

\begin{figure}[t]
    \centering
    \includegraphics[width=\textwidth]{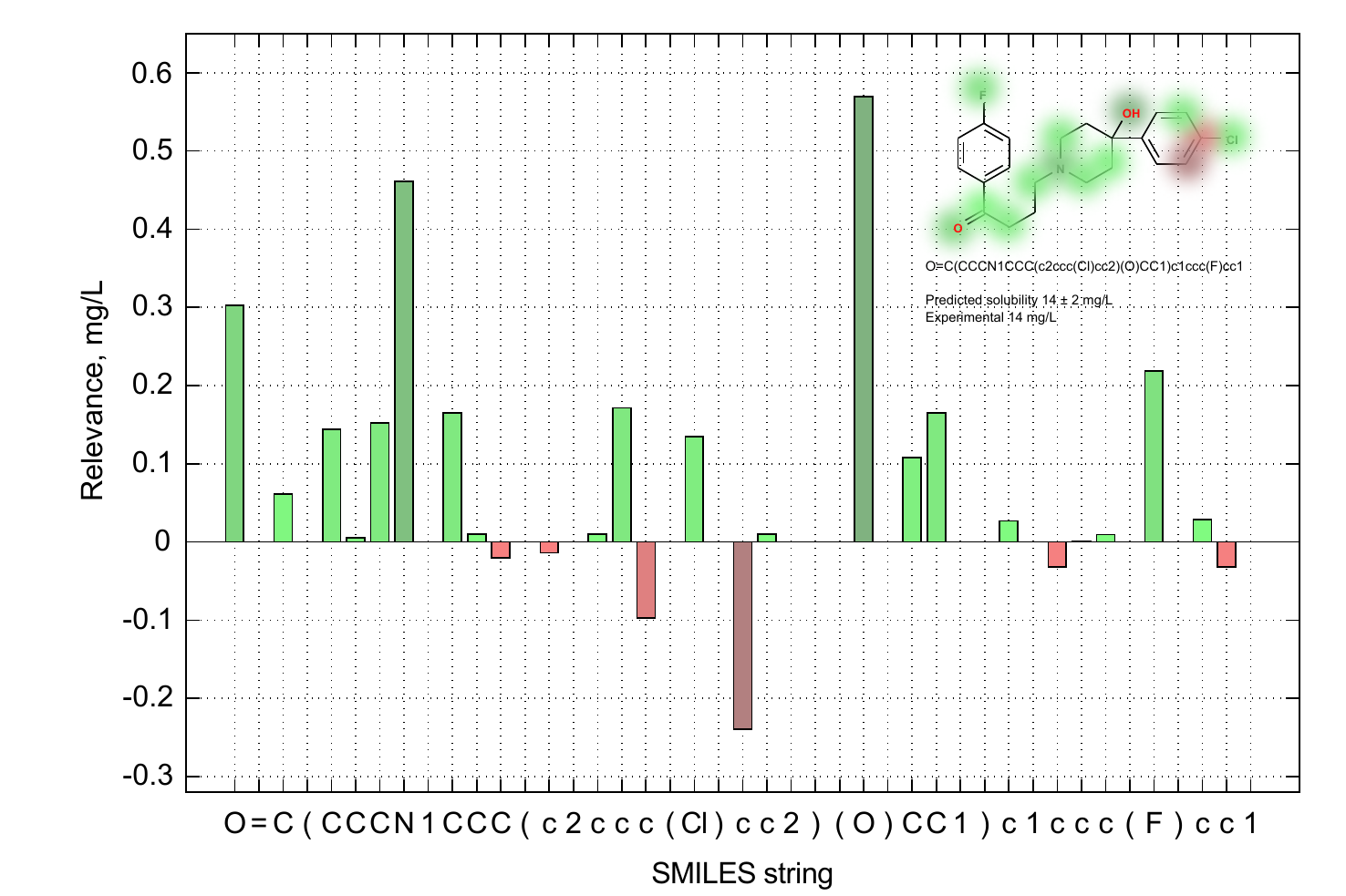}
    \caption{Visualization of atom contributions to aqueous solubility of haloperidol. The greep bars stand for more soluble features, whereas the red ones show the opposite effect.}
    \label{fig:sol}
\end{figure}

The Transformer model calculated the same solubility 14 $\pm$ 2 mg/L for this compound. The individual atom contributions are shown in Fig.~\ref{fig:sol}. Hydroxyl, carbonyl, aliphatic nitrogen, and halogens contributed mostly to the solubility. These groups can form ionizable zones in the molecule thus helping water to dissolve the substance. Several aromatic carbons had negative contributions, which was expected since aromatic compounds are poorly soluble in water. Thus the overall explanation made sense, and the model had an excellent statistics not because of spurious correlations, but because it found the right fragmental features responsible for modelled property. The standalone program contributed in this work has no dependencies on machine learning frameworks, it is easy to install, to use, and to interpret the modelling results. This will make it an indispensable work-horse for drug-development projects world-wide.

\section{Conclusions and outlook}

For the first time we proposed a SMILES canonicalization method based on Transformer architecture that extracts information-rich real-value embeddings during the encoding process and exposes them for further QSAR studies. Also, for the first time we developed a framework for the interpretation of models based on the Transformer architecture using a layer-wise relevance propagation (LPR) approach.

TextCNN approaches efficiently worked with embeddings generated by Transformer, and the final quality of the QSAR models was higher compared to the models obtained with the state-of-the-art methods on the majority of diverse benchmark datasets. The Transformer-CNN architecture required less than a hundred iterations to converge for QSAR tasks to model various biological activity or physico-chemical properties. It can be easily embedded into de-novo drug development pipelines. The model predictions interpreted in a fragment contribution manner using the LPR could be useful to design new molecules with desired biological activity and ADMETox properties. The source code is available on \url{https://github.com/bigchem/transformer-cnn} as well as an on-line version on \url{http://ochem.eu}. For solubility and AMES mutagenicity we also deposited standalone models in the GitHub repository, which not only predict the respective properties but also provide interpretations of predictions.

The Transformer-CNN predicts the endpoint based on an average of individual prognosis for a batch of augmented SMILES belonging to the same molecule. The deviation within the batch can serve as a measure of a confidence interval of the prognosis. Dissipation of relevance on biases as well as analysis of restored SMILES can be used to derive the applicability domains of models. These questions will be addressed in the upcoming studies.

Also, as a comment, we do not think that the authors benchmarking their methods are impassioned about their work. Such benchmarking could be properly done by other users, and we do hope to see the proposed method used soon in future publications. But indeed, remarkably, in this work we saw an outstanding performance of the proposed architecture, which provided systematically better or at least similar results compared to the best descriptor-based approaches as well as several analysed deep neural network architectures. Even more remarkably, the Transformer CNN has practically no adjustable meta parameters and thus does not require spending time to tune hyperparameters of neural architectures, use the grid search to optimise Support Vector Machines, optimise multiple parameters of XGBoost, apply various descriptors filtering and preprocessing, which could easily contribute to the overfitting of models. This as well as the possibility to interpret models makes Transformer CNN a Swiss-knife for QSAR modeling and interpretation.

\section{Acknowledgement}

The project leading to this report has received funding from the European Union’s Horizon 2020 research and innovation program under the Marie Skłodowska-Curie grant agreement No 676434, “Big Data in Chemistry” and ERA-CVD "CardioOncology" project, BMBF 01KL1710. The authors thank NVIDIA Corporation for donating Quadro P6000, Titan Xp, and Titan V graphics cards for this research work. 
 
The article reflects only the author’s view and neither the European Commission nor the Research Executive Agency (REA) are responsible for any use that may be made of the information it contains.

\bibliographystyle{naturemag}  
\bibliography{references}

\begin{thebibliography}{10}
\expandafter\ifx\csname url\endcsname\relax
  \def\url#1{\texttt{#1}}\fi
\expandafter\ifx\csname urlprefix\endcsname\relax\def\urlprefix{URL }\fi
\providecommand{\bibinfo}[2]{#2}
\providecommand{\eprint}[2][]{\url{#2}}

\bibitem{AttentionArticle}
\bibinfo{author}{{Vaswani}, A.} \emph{et~al.}
\newblock \bibinfo{title}{{Attention Is All You Need}}.
\newblock \emph{\bibinfo{journal}{arXiv e-prints}}
  \bibinfo{pages}{arXiv:1706.03762} (\bibinfo{year}{2017}).
\newblock \eprint{1706.03762}.

\bibitem{CharNN}
\bibinfo{author}{{Zhang}, X.}, \bibinfo{author}{{Zhao}, J.} \&
  \bibinfo{author}{{LeCun}, Y.}
\newblock \bibinfo{title}{{Character-level Convolutional Networks for Text
  Classification}}.
\newblock \emph{\bibinfo{journal}{arXiv e-prints}}
  \bibinfo{pages}{arXiv:1509.01626} (\bibinfo{year}{2015}).
\newblock \eprint{1509.01626}.

\bibitem{OCHEM}
\bibinfo{author}{{Sushko}, I.} \emph{et~al.}
\newblock \bibinfo{title}{{Online chemical modeling environment (OCHEM): web
  platform for data storage, model development and publishing of chemical
  information}}.
\newblock \emph{\bibinfo{journal}{Journal of Computer-Aided Molecular Design}}
  \textbf{\bibinfo{volume}{25}}, \bibinfo{pages}{533--554}
  (\bibinfo{year}{2011}).

\bibitem{Todeschini}
\bibinfo{author}{Mauri, A.}, \bibinfo{author}{Consonni, V.},
  \bibinfo{author}{Pavan, M.} \& \bibinfo{author}{Todeschini, R.}
\newblock \bibinfo{title}{Dragon software: An easy approach to molecular
  descriptor calculations}.
\newblock \emph{\bibinfo{journal}{MATCH Communications in Mathematical and in
  Computer Chemistry}} \textbf{\bibinfo{volume}{56}}, \bibinfo{pages}{237--248}
  (\bibinfo{year}{2006}).

\bibitem{FragmentDescriptors}
\bibinfo{author}{Baskin, I.}
\newblock \bibinfo{title}{Chapter 1 fragment descriptors in sar/qsar/qspr
  studies{,} molecular similarity analysis and in virtual screening}.
\newblock In \bibinfo{editor}{Varnek, A.} (ed.)
  \emph{\bibinfo{booktitle}{Chemoinformatics Approaches to Virtual Screening}},
  \bibinfo{pages}{1--43} (\bibinfo{publisher}{The Royal Society of Chemistry},
  \bibinfo{year}{2008}).
\newblock \urlprefix\url{http://dx.doi.org/10.1039/9781847558879-00001}.

\bibitem{FeatureSelection}
\bibinfo{author}{Eklund, M.}, \bibinfo{author}{Norinder, U.},
  \bibinfo{author}{Boyer, S.} \& \bibinfo{author}{Carlsson, L.}
\newblock \bibinfo{title}{Choosing feature selection and learning algorithms in
  qsar}.
\newblock \emph{\bibinfo{journal}{Journal of Chemical Information and
  Modeling}} \textbf{\bibinfo{volume}{54}}, \bibinfo{pages}{837--843}
  (\bibinfo{year}{2014}).
\newblock \urlprefix\url{https://doi.org/10.1021/ci400573c}.
\newblock \bibinfo{note}{PMID: 24460242},
  \eprint{https://doi.org/10.1021/ci400573c}.

\bibitem{RiseOfDeepLearning}
\bibinfo{author}{Baskin, I.~I.}, \bibinfo{author}{Winkler, D.} \&
  \bibinfo{author}{Tetko, I.~V.}
\newblock \bibinfo{title}{A renaissance of neural networks in drug discovery}.
\newblock \emph{\bibinfo{journal}{Expert Opinion on Drug Discovery}}
  \textbf{\bibinfo{volume}{11}}, \bibinfo{pages}{785--795}
  (\bibinfo{year}{2016}).
\newblock \urlprefix\url{https://doi.org/10.1080/17460441.2016.1201262}.
\newblock \bibinfo{note}{PMID: 27295548},
  \eprint{https://doi.org/10.1080/17460441.2016.1201262}.

\bibitem{NeurlaFingerprint}
\bibinfo{author}{{Duvenaud}, D.} \emph{et~al.}
\newblock \bibinfo{title}{{Convolutional Networks on Graphs for Learning
  Molecular Fingerprints}}.
\newblock \emph{\bibinfo{journal}{arXiv e-prints}}
  \bibinfo{pages}{arXiv:1509.09292} (\bibinfo{year}{2015}).
\newblock \eprint{1509.09292}.

\bibitem{Coley}
\bibinfo{author}{Coley, C.~W.}, \bibinfo{author}{Barzilay, R.},
  \bibinfo{author}{Green, W.~H.}, \bibinfo{author}{Jaakkola, T.~S.} \&
  \bibinfo{author}{Jensen, K.~F.}
\newblock \bibinfo{title}{Convolutional embedding of attributed molecular
  graphs for physical property prediction}.
\newblock \emph{\bibinfo{journal}{Journal of Chemical Information and
  Modeling}} \textbf{\bibinfo{volume}{57}}, \bibinfo{pages}{1757--1772}
  (\bibinfo{year}{2017}).
\newblock \urlprefix\url{https://doi.org/10.1021/acs.jcim.6b00601}.
\newblock \bibinfo{note}{PMID: 28696688},
  \eprint{https://doi.org/10.1021/acs.jcim.6b00601}.

\bibitem{Bombarelli}
\bibinfo{author}{Gómez-Bombarelli, R.} \emph{et~al.}
\newblock \bibinfo{title}{Automatic chemical design using a data-driven
  continuous representation of molecules}.
\newblock \emph{\bibinfo{journal}{ACS Central Science}}
  \textbf{\bibinfo{volume}{4}}, \bibinfo{pages}{268--276}
  (\bibinfo{year}{2018}).
\newblock \urlprefix\url{https://doi.org/10.1021/acscentsci.7b00572}.
\newblock \bibinfo{note}{PMID: 29532027},
  \eprint{https://doi.org/10.1021/acscentsci.7b00572}.

\bibitem{Augmentation}
\bibinfo{author}{{Kimber}, T.~B.}, \bibinfo{author}{{Engelke}, S.},
  \bibinfo{author}{{Tetko}, I.~V.}, \bibinfo{author}{{Bruno}, E.} \&
  \bibinfo{author}{{Godin}, G.}
\newblock \bibinfo{title}{{Synergy Effect between Convolutional Neural Networks
  and the Multiplicity of SMILES for Improvement of Molecular Prediction}}.
\newblock \emph{\bibinfo{journal}{arXiv e-prints}}
  \bibinfo{pages}{arXiv:1812.04439} (\bibinfo{year}{2018}).
\newblock \eprint{1812.04439}.

\bibitem{MessagePassing}
\bibinfo{author}{{Gilmer}, J.}, \bibinfo{author}{{Schoenholz}, S.~S.},
  \bibinfo{author}{{Riley}, P.~F.}, \bibinfo{author}{{Vinyals}, O.} \&
  \bibinfo{author}{{Dahl}, G.~E.}
\newblock \bibinfo{title}{{Neural Message Passing for Quantum Chemistry}}.
\newblock \emph{\bibinfo{journal}{arXiv e-prints}}
  \bibinfo{pages}{arXiv:1704.01212} (\bibinfo{year}{2017}).
\newblock \eprint{1704.01212}.

\bibitem{EdgeAttention}
\bibinfo{author}{{Shang}, C.} \emph{et~al.}
\newblock \bibinfo{title}{{Edge Attention-based Multi-Relational Graph
  Convolutional Networks}}.
\newblock \emph{\bibinfo{journal}{arXiv e-prints}}
  \bibinfo{pages}{arXiv:1802.04944} (\bibinfo{year}{2018}).
\newblock \eprint{1802.04944}.

\bibitem{Weininger}
\bibinfo{author}{Weininger, D.}
\newblock \bibinfo{title}{Smiles, a chemical language and information system.
  1. introduction to methodology and encoding rules}.
\newblock \emph{\bibinfo{journal}{Journal of Chemical Information and Computer
  Sciences}} \textbf{\bibinfo{volume}{28}}, \bibinfo{pages}{31--36}
  (\bibinfo{year}{1988}).
\newblock \urlprefix\url{https://pubs.acs.org/doi/abs/10.1021/ci00057a005}.
\newblock \eprint{https://pubs.acs.org/doi/pdf/10.1021/ci00057a005}.

\bibitem{OpenBabel}
\bibinfo{author}{O'Boyle, N.~M.} \emph{et~al.}
\newblock \bibinfo{title}{Open babel: An open chemical toolbox}.
\newblock \emph{\bibinfo{journal}{Journal of Cheminformatics}}
  \textbf{\bibinfo{volume}{3}}, \bibinfo{pages}{33} (\bibinfo{year}{2011}).
\newblock \urlprefix\url{https://doi.org/10.1186/1758-2946-3-33}.

\bibitem{LINGO}
\bibinfo{author}{Vidal, D.}, \bibinfo{author}{Thormann, M.} \&
  \bibinfo{author}{Pons, M.}
\newblock \bibinfo{title}{Lingo, an efficient holographic text based method to
  calculate biophysical properties and intermolecular similarities}.
\newblock \emph{\bibinfo{journal}{Journal of Chemical Information and
  Modeling}} \textbf{\bibinfo{volume}{45}}, \bibinfo{pages}{386--393}
  (\bibinfo{year}{2005}).
\newblock \urlprefix\url{https://doi.org/10.1021/ci0496797}.
\newblock \bibinfo{note}{PMID: 15807504},
  \eprint{https://doi.org/10.1021/ci0496797}.

\bibitem{LeCunCharNN}
\bibinfo{author}{{Zhang}, X.} \& \bibinfo{author}{{LeCun}, Y.}
\newblock \bibinfo{title}{{Text Understanding from Scratch}}.
\newblock \emph{\bibinfo{journal}{arXiv e-prints}}
  \bibinfo{pages}{arXiv:1502.01710} (\bibinfo{year}{2015}).
\newblock \eprint{1502.01710}.

\bibitem{Smiles2Vec}
\bibinfo{author}{{Goh}, G.~B.}, \bibinfo{author}{{Hodas}, N.~O.},
  \bibinfo{author}{{Siegel}, C.} \& \bibinfo{author}{{Vishnu}, A.}
\newblock \bibinfo{title}{{SMILES2Vec: An Interpretable General-Purpose Deep
  Neural Network for Predicting Chemical Properties}}.
\newblock \emph{\bibinfo{journal}{arXiv e-prints}}
  \bibinfo{pages}{arXiv:1712.02034} (\bibinfo{year}{2017}).
\newblock \eprint{1712.02034}.

\bibitem{LearningSmiles}
\bibinfo{author}{{Jastrz{\k{e}}bski}, S.}, \bibinfo{author}{{Le{\'s}niak}, D.}
  \& \bibinfo{author}{{Czarnecki}, W.~M.}
\newblock \bibinfo{title}{{Learning to SMILE(S)}}.
\newblock \emph{\bibinfo{journal}{arXiv e-prints}}
  \bibinfo{pages}{arXiv:1602.06289} (\bibinfo{year}{2016}).
\newblock \eprint{1602.06289}.

\bibitem{chemNet}
\bibinfo{author}{{Goh}, G.~B.}, \bibinfo{author}{{Siegel}, C.},
  \bibinfo{author}{{Vishnu}, A.} \& \bibinfo{author}{{Hodas}, N.~O.}
\newblock \bibinfo{title}{{Using Rule-Based Labels for Weak Supervised
  Learning: A ChemNet for Transferable Chemical Property Prediction}}.
\newblock \emph{\bibinfo{journal}{arXiv e-prints}}
  \bibinfo{pages}{arXiv:1712.02734} (\bibinfo{year}{2017}).
\newblock \eprint{1712.02734}.

\bibitem{SmilesAttention}
\bibinfo{author}{Zheng, S.}, \bibinfo{author}{Yan, X.}, \bibinfo{author}{Yang,
  Y.} \& \bibinfo{author}{Xu, J.}
\newblock \bibinfo{title}{Identifying structure-property relationships through
  smiles syntax analysis with self-attention mechanism}.
\newblock \emph{\bibinfo{journal}{Journal of Chemical Information and
  Modeling}} \textbf{\bibinfo{volume}{59}}, \bibinfo{pages}{914--923}
  (\bibinfo{year}{2019}).
\newblock \urlprefix\url{https://doi.org/10.1021/acs.jcim.8b00803}.
\newblock \bibinfo{note}{PMID: 30669836},
  \eprint{https://doi.org/10.1021/acs.jcim.8b00803}.

\bibitem{TetkoAugmentation}
\bibinfo{author}{Tetko, I.~V.}, \bibinfo{author}{Karpov, P.},
  \bibinfo{author}{Bruno, E.}, \bibinfo{author}{Kimber, T.~B.} \&
  \bibinfo{author}{Godin, G.}
\newblock \bibinfo{title}{Augmentation is what you need!}
\newblock In \bibinfo{editor}{Tetko, I.~V.},
  \bibinfo{editor}{K{\r{u}}rkov{\'a}, V.}, \bibinfo{editor}{Karpov, P.} \&
  \bibinfo{editor}{Theis, F.} (eds.) \emph{\bibinfo{booktitle}{Artificial
  Neural Networks and Machine Learning -- ICANN 2019: Workshop and Special
  Sessions}}, \bibinfo{pages}{831--835} (\bibinfo{publisher}{Springer
  International Publishing}, \bibinfo{address}{Cham}, \bibinfo{year}{2019}).

\bibitem{ImageEmbeddings}
\bibinfo{author}{Kiela, D.} \& \bibinfo{author}{Bottou, L.}
\newblock \bibinfo{title}{Learning image embeddings using convolutional neural
  networks for improved multi-modal semantics}.
\newblock In \emph{\bibinfo{booktitle}{Proceedings of the 2014 Conference on
  Empirical Methods in Natural Language Processing ({EMNLP})}},
  \bibinfo{pages}{36--45} (\bibinfo{publisher}{Association for Computational
  Linguistics}, \bibinfo{address}{Doha, Qatar}, \bibinfo{year}{2014}).
\newblock \urlprefix\url{https://www.aclweb.org/anthology/D14-1005}.

\bibitem{TextEmbeddings}
\bibinfo{author}{Pennington, J.}, \bibinfo{author}{Socher, R.} \&
  \bibinfo{author}{Manning, C.}
\newblock \bibinfo{title}{{G}love: Global vectors for word representation}.
\newblock In \emph{\bibinfo{booktitle}{Proceedings of the 2014 Conference on
  Empirical Methods in Natural Language Processing ({EMNLP})}},
  \bibinfo{pages}{1532--1543} (\bibinfo{publisher}{Association for
  Computational Linguistics}, \bibinfo{address}{Doha, Qatar},
  \bibinfo{year}{2014}).
\newblock \urlprefix\url{https://www.aclweb.org/anthology/D14-1162}.

\bibitem{Autoencoders}
\bibinfo{author}{Hinton, G.~E.} \& \bibinfo{author}{Salakhutdinov, R.~R.}
\newblock \bibinfo{title}{Reducing the dimensionality of data with neural
  networks}.
\newblock \emph{\bibinfo{journal}{Science}} \textbf{\bibinfo{volume}{313}},
  \bibinfo{pages}{504--507} (\bibinfo{year}{2006}).
\newblock \urlprefix\url{https://science.sciencemag.org/content/313/5786/504}.
\newblock
  \eprint{https://science.sciencemag.org/content/313/5786/504.full.pdf}.

\bibitem{InChi}
\bibinfo{author}{Heller, S.}, \bibinfo{author}{McNaught, A.},
  \bibinfo{author}{Stein, S.}, \bibinfo{author}{Tchekhovskoi, D.} \&
  \bibinfo{author}{Pletnev, I.}
\newblock \bibinfo{title}{Inchi - the worldwide chemical structure identifier
  standard}.
\newblock \emph{\bibinfo{journal}{Journal of Cheminformatics}}
  \textbf{\bibinfo{volume}{5}}, \bibinfo{pages}{7} (\bibinfo{year}{2013}).
\newblock \urlprefix\url{https://doi.org/10.1186/1758-2946-5-7}.

\bibitem{smi2Inchi}
\bibinfo{author}{Winter, R.}, \bibinfo{author}{Montanari, F.},
  \bibinfo{author}{Noé, F.} \& \bibinfo{author}{Clevert, D.-A.}
\newblock \bibinfo{title}{Learning continuous and data-driven molecular
  descriptors by translating equivalent chemical representations}.
\newblock \emph{\bibinfo{journal}{Chem. Sci.}} \textbf{\bibinfo{volume}{10}},
  \bibinfo{pages}{1692--1701} (\bibinfo{year}{2019}).
\newblock \urlprefix\url{http://dx.doi.org/10.1039/C8SC04175J}.

\bibitem{LSTM}
\bibinfo{author}{Hochreiter, S.} \& \bibinfo{author}{Schmidhuber, J.}
\newblock \bibinfo{title}{Long short-term memory}.
\newblock \emph{\bibinfo{journal}{Neural Comput.}}
  \textbf{\bibinfo{volume}{9}}, \bibinfo{pages}{1735–1780}
  (\bibinfo{year}{1997}).
\newblock \urlprefix\url{https://doi.org/10.1162/neco.1997.9.8.1735}.

\bibitem{Schwaller}
\bibinfo{author}{Schwaller, P.} \emph{et~al.}
\newblock \bibinfo{title}{Molecular transformer: A model for
  uncertainty-calibrated chemical reaction prediction}.
\newblock \emph{\bibinfo{journal}{ACS Central Science}}
  \textbf{\bibinfo{volume}{5}}, \bibinfo{pages}{1572--1583}
  (\bibinfo{year}{2019}).
\newblock \urlprefix\url{https://doi.org/10.1021/acscentsci.9b00576}.
\newblock \eprint{https://doi.org/10.1021/acscentsci.9b00576}.

\bibitem{KarpovRetro}
\bibinfo{author}{Karpov, P.}, \bibinfo{author}{Godin, G.} \&
  \bibinfo{author}{Tetko, I.~V.}
\newblock \bibinfo{title}{A transformer model for retrosynthesis}.
\newblock In \bibinfo{editor}{Tetko, I.~V.},
  \bibinfo{editor}{K{\r{u}}rkov{\'a}, V.}, \bibinfo{editor}{Karpov, P.} \&
  \bibinfo{editor}{Theis, F.} (eds.) \emph{\bibinfo{booktitle}{Artificial
  Neural Networks and Machine Learning -- ICANN 2019: Workshop and Special
  Sessions}}, \bibinfo{pages}{817--830} (\bibinfo{publisher}{Springer
  International Publishing}, \bibinfo{address}{Cham}, \bibinfo{year}{2019}).

\bibitem{Samek2019}
\bibinfo{author}{Samek, W.} \& \bibinfo{author}{M{\"u}ller, K.-R.}
\newblock \emph{\bibinfo{title}{Towards Explainable Artificial Intelligence}},
  \bibinfo{pages}{5--22} (\bibinfo{publisher}{Springer International
  Publishing}, \bibinfo{address}{Cham}, \bibinfo{year}{2019}).
\newblock \urlprefix\url{https://doi.org/10.1007/978-3-030-28954-6_1}.

\bibitem{Montavon}
\bibinfo{author}{Montavon, G.}, \bibinfo{author}{Binder, A.},
  \bibinfo{author}{Lapuschkin, S.}, \bibinfo{author}{Samek, W.} \&
  \bibinfo{author}{M{\"u}ller, K.-R.}
\newblock \emph{\bibinfo{title}{Layer-Wise Relevance Propagation: An
  Overview}}, \bibinfo{pages}{193--209} (\bibinfo{publisher}{Springer
  International Publishing}, \bibinfo{address}{Cham}, \bibinfo{year}{2019}).
\newblock \urlprefix\url{https://doi.org/10.1007/978-3-030-28954-6_10}.

\bibitem{TetkoPruning}
\bibinfo{author}{Tetko, I.~V.}, \bibinfo{author}{Villa, A. E.~P.} \&
  \bibinfo{author}{Livingstone, D.~J.}
\newblock \bibinfo{title}{Neural network studies. 2. variable selection}.
\newblock \emph{\bibinfo{journal}{Journal of Chemical Information and Computer
  Sciences}} \textbf{\bibinfo{volume}{36}}, \bibinfo{pages}{794--803}
  (\bibinfo{year}{1996}).
\newblock \urlprefix\url{https://doi.org/10.1021/ci950204c}.
\newblock \bibinfo{note}{PMID: 8768768},
  \eprint{https://doi.org/10.1021/ci950204c}.

\bibitem{ChEMBL}
\bibinfo{author}{Gaulton, A.} \emph{et~al.}
\newblock \bibinfo{title}{{ChEMBL: a large-scale bioactivity database for drug
  discovery}}.
\newblock \emph{\bibinfo{journal}{Nucleic Acids Research}}
  \textbf{\bibinfo{volume}{40}}, \bibinfo{pages}{D1100--D1107}
  (\bibinfo{year}{2011}).
\newblock \urlprefix\url{https://doi.org/10.1093/nar/gkr777}.
\newblock
  \eprint{https://academic.oup.com/nar/article-pdf/40/D1/D1100/16955876/gkr777.pdf}.

\bibitem{Segler}
\bibinfo{author}{Segler, M. H.~S.}, \bibinfo{author}{Kogej, T.},
  \bibinfo{author}{Tyrchan, C.} \& \bibinfo{author}{Waller, M.~P.}
\newblock \bibinfo{title}{Generating focused molecule libraries for drug
  discovery with recurrent neural networks}.
\newblock \emph{\bibinfo{journal}{ACS Central Science}}
  \textbf{\bibinfo{volume}{4}}, \bibinfo{pages}{120--131}
  (\bibinfo{year}{2018}).
\newblock \urlprefix\url{https://doi.org/10.1021/acscentsci.7b00512}.
\newblock \bibinfo{note}{PMID: 29392184},
  \eprint{https://doi.org/10.1021/acscentsci.7b00512}.

\bibitem{Gupta}
\bibinfo{author}{Gupta, A.} \emph{et~al.}
\newblock \bibinfo{title}{Generative recurrent networks for de novo drug
  design}.
\newblock \emph{\bibinfo{journal}{Molecular Informatics}}
  \textbf{\bibinfo{volume}{37}}, \bibinfo{pages}{1700111}
  (\bibinfo{year}{2018}).
\newblock
  \urlprefix\url{https://onlinelibrary.wiley.com/doi/abs/10.1002/minf.201700111}.
\newblock
  \eprint{https://onlinelibrary.wiley.com/doi/pdf/10.1002/minf.201700111}.

\bibitem{AnnotatedTransformer}
\bibinfo{author}{Rush, A.}
\newblock \bibinfo{title}{The annotated transformer}.
\newblock In \emph{\bibinfo{booktitle}{Proceedings of Workshop for {NLP} Open
  Source Software ({NLP}-{OSS})}}, \bibinfo{pages}{52--60}
  (\bibinfo{publisher}{Association for Computational Linguistics},
  \bibinfo{address}{Melbourne, Australia}, \bibinfo{year}{2018}).
\newblock \urlprefix\url{https://www.aclweb.org/anthology/W18-2509}.

\bibitem{tensorflow}
\bibinfo{author}{Abadi, M.} \emph{et~al.}
\newblock \bibinfo{title}{{TensorFlow}: Large-scale machine learning on
  heterogeneous systems} (\bibinfo{year}{2015}).
\newblock \urlprefix\url{http://tensorflow.org/}.
\newblock \bibinfo{note}{Software available from tensorflow.org}.

\bibitem{rdkit}
\bibinfo{author}{Landrum, G.}
\newblock \bibinfo{title}{Rdkit: Open-source cheminformatics}.
\newblock \urlprefix\url{http://www.rdkit.org}.

\bibitem{deepchem}
\bibinfo{author}{Ramsundar, B.}, \bibinfo{author}{Eastman, P.},
  \bibinfo{author}{Walters, P.} \& \bibinfo{author}{Pande, V.}
\newblock \emph{\bibinfo{title}{Deep Learning for the Life Sciences: Applying
  Deep Learning to Genomics, Microscopy, Drug Discovery, and More}}
  (\bibinfo{publisher}{``O'Reilly Media, Inc.''}, \bibinfo{year}{2019}).

\bibitem{dropout}
\bibinfo{author}{Srivastava, N.}, \bibinfo{author}{Hinton, G.},
  \bibinfo{author}{Krizhevsky, A.}, \bibinfo{author}{Sutskever, I.} \&
  \bibinfo{author}{Salakhutdinov, R.}
\newblock \bibinfo{title}{Dropout: A simple way to prevent neural networks from
  overfitting}.
\newblock \emph{\bibinfo{journal}{Journal of Machine Learning Research}}
  \textbf{\bibinfo{volume}{15}}, \bibinfo{pages}{1929--1958}
  (\bibinfo{year}{2014}).
\newblock \urlprefix\url{http://jmlr.org/papers/v15/srivastava14a.html}.

\bibitem{highway}
\bibinfo{author}{{Srivastava}, R.~K.}, \bibinfo{author}{{Greff}, K.} \&
  \bibinfo{author}{{Schmidhuber}, J.}
\newblock \bibinfo{title}{{Highway Networks}}.
\newblock \emph{\bibinfo{journal}{arXiv e-prints}}
  \bibinfo{pages}{arXiv:1505.00387} (\bibinfo{year}{2015}).
\newblock \eprint{1505.00387}.

\bibitem{Draper}
\bibinfo{author}{Draper, N.~R.} \& \bibinfo{author}{Smith, H.}
\newblock \emph{\bibinfo{title}{Applied Regression Analysis}}
  (\bibinfo{publisher}{John Wiley \& Sons}, \bibinfo{year}{2014}).

\bibitem{MP}
\bibinfo{author}{Tetko, I.~V.} \emph{et~al.}
\newblock \bibinfo{title}{How accurately can we predict the melting points of
  drug-like compounds?}
\newblock \emph{\bibinfo{journal}{Journal of Chemical Information and
  Modeling}} \textbf{\bibinfo{volume}{54}}, \bibinfo{pages}{3320--3329}
  (\bibinfo{year}{2014}).
\newblock \urlprefix\url{https://doi.org/10.1021/ci5005288}.
\newblock \bibinfo{note}{PMID: 25489863},
  \eprint{https://doi.org/10.1021/ci5005288}.

\bibitem{benchmark}
\bibinfo{author}{Wu, Z.} \emph{et~al.}
\newblock \bibinfo{title}{Moleculenet: a benchmark for molecular machine
  learning}.
\newblock \emph{\bibinfo{journal}{Chem. Sci.}} \textbf{\bibinfo{volume}{9}},
  \bibinfo{pages}{513--530} (\bibinfo{year}{2018}).
\newblock \urlprefix\url{http://dx.doi.org/10.1039/C7SC02664A}.

\bibitem{BP}
\bibinfo{author}{Brandmaier, S.}, \bibinfo{author}{Sahlin, U.},
  \bibinfo{author}{Tetko, I.~V.} \& \bibinfo{author}{{\"O}berg, T.}
\newblock \bibinfo{title}{{PLS-Optimal}: A stepwise {D-Optimal} design based on
  latent variables}.
\newblock \emph{\bibinfo{journal}{Journal of Chemical Information and
  Modeling}} \textbf{\bibinfo{volume}{52}}, \bibinfo{pages}{975--983}
  (\bibinfo{year}{2012}).

\bibitem{ames}
\bibinfo{author}{Sushko, I.} \emph{et~al.}
\newblock \bibinfo{title}{Applicability domains for classification problems:
  Benchmarking of distance to models for ames mutagenicity set}.
\newblock \emph{\bibinfo{journal}{Journal of Chemical Information and
  Modeling}} \textbf{\bibinfo{volume}{50}}, \bibinfo{pages}{2094--2111}
  (\bibinfo{year}{2010}).
\newblock \urlprefix\url{https://doi.org/10.1021/ci100253r}.
\newblock \bibinfo{note}{PMID: 21033656},
  \eprint{https://doi.org/10.1021/ci100253r}.

\bibitem{Tetkosolubility}
\bibinfo{author}{Tetko, I.~V.}, \bibinfo{author}{Tanchuk, V.~Y.},
  \bibinfo{author}{Kasheva, T.~N.} \& \bibinfo{author}{Villa, A. E.~P.}
\newblock \bibinfo{title}{Estimation of aqueous solubility of chemical
  compounds using e-state indices}.
\newblock \emph{\bibinfo{journal}{Journal of Chemical Information and Computer
  Sciences}} \textbf{\bibinfo{volume}{41}}, \bibinfo{pages}{1488--1493}
  (\bibinfo{year}{2001}).
\newblock \urlprefix\url{https://doi.org/10.1021/ci000392t}.
\newblock \bibinfo{note}{PMID: 11749573},
  \eprint{https://doi.org/10.1021/ci000392t}.

\bibitem{TetkoLipo}
\bibinfo{author}{Huuskonen, J.~J.}, \bibinfo{author}{Livingstone, D.~J.} \&
  \bibinfo{author}{Tetko, I.~V.}
\newblock \bibinfo{title}{Neural network modeling for estimation of partition
  coefficient based on atom-type electrotopological state indices}.
\newblock \emph{\bibinfo{journal}{Journal of Chemical Information and Computer
  Sciences}} \textbf{\bibinfo{volume}{40}}, \bibinfo{pages}{947--955}
  (\bibinfo{year}{2000}).
\newblock \urlprefix\url{https://doi.org/10.1021/ci9904261}.
\newblock \bibinfo{note}{PMID: 10955523},
  \eprint{https://doi.org/10.1021/ci9904261}.

\bibitem{jak3}
\bibinfo{author}{Suzuki, K.} \emph{et~al.}
\newblock \bibinfo{title}{{Janus kinase 3 (Jak3) is essential for common
  cytokine receptor $\gamma$ chain ($\gamma$c)-dependent signaling: comparative
  analysis of $\gamma$c, Jak3, and $\gamma$c and Jak3 double-deficient mice}}.
\newblock \emph{\bibinfo{journal}{International Immunology}}
  \textbf{\bibinfo{volume}{12}}, \bibinfo{pages}{123--132}
  (\bibinfo{year}{2000}).
\newblock \urlprefix\url{https://doi.org/10.1093/intimm/12.2.123}.
\newblock
  \eprint{https://academic.oup.com/intimm/article-pdf/12/2/123/18318734/123.pdf}.

\bibitem{dhfr}
\bibinfo{author}{Sutherland, J.~J.} \& \bibinfo{author}{Weaver, D.~F.}
\newblock \bibinfo{title}{Three-dimensional quantitative structure-activity and
  structure-selectivity relationships of dihydrofolate reductase inhibitors}.
\newblock \emph{\bibinfo{journal}{Journal of Computer-Aided Molecular Design}}
  \textbf{\bibinfo{volume}{18}}, \bibinfo{pages}{309--331}
  (\bibinfo{year}{2004}).

\bibitem{biodeg}
\bibinfo{author}{Vorberg, S.} \& \bibinfo{author}{Tetko, I.~V.}
\newblock \bibinfo{title}{Modeling the biodegradability of chemical compounds
  using the online {CHEmical} modeling environment ({OCHEM})}.
\newblock \emph{\bibinfo{journal}{Mol. Inform.}} \textbf{\bibinfo{volume}{33}},
  \bibinfo{pages}{73--85} (\bibinfo{year}{2014}).

\bibitem{lel}
\bibinfo{author}{Novotarskyi, S.} \emph{et~al.}
\newblock \bibinfo{title}{Toxcast epa in vitro to in vivo challenge: Insight
  into the rank-i model}.
\newblock \emph{\bibinfo{journal}{Chemical Research in Toxicology}}
  \textbf{\bibinfo{volume}{29}}, \bibinfo{pages}{768--775}
  (\bibinfo{year}{2016}).
\newblock \urlprefix\url{https://doi.org/10.1021/acs.chemrestox.5b00481}.
\newblock \bibinfo{note}{PMID: 27120770},
  \eprint{https://doi.org/10.1021/acs.chemrestox.5b00481}.

\bibitem{rpar}
\bibinfo{author}{Rybacka, A.}, \bibinfo{author}{Rudén, C.},
  \bibinfo{author}{Tetko, I.~V.} \& \bibinfo{author}{Andersson, P.~L.}
\newblock \bibinfo{title}{Identifying potential endocrine disruptors among
  industrial chemicals and their metabolites – development and evaluation of
  in silico tools}.
\newblock \emph{\bibinfo{journal}{Chemosphere}} \textbf{\bibinfo{volume}{139}},
  \bibinfo{pages}{372 -- 378} (\bibinfo{year}{2015}).
\newblock
  \urlprefix\url{http://www.sciencedirect.com/science/article/pii/S0045653515007560}.

\bibitem{xia}
\bibinfo{author}{Xia, Z.}, \bibinfo{author}{Karpov, P.},
  \bibinfo{author}{Popowicz, G.} \& \bibinfo{author}{Tetko, I.~V.}
\newblock \bibinfo{title}{Focused library generator: case of mdmx inhibitors}.
\newblock \emph{\bibinfo{journal}{Journal of Computer-Aided Molecular Design}}
  (\bibinfo{year}{2019}).
\newblock \urlprefix\url{https://doi.org/10.1007/s10822-019-00242-8}.

\bibitem{libsvm}
\bibinfo{author}{Chang, C.-C.} \& \bibinfo{author}{Lin, C.-J.}
\newblock \bibinfo{title}{Libsvm: A library for support vector machines}.
\newblock \emph{\bibinfo{journal}{ACM Trans. Intell. Syst. Technol.}}
  \textbf{\bibinfo{volume}{2}} (\bibinfo{year}{2011}).
\newblock \urlprefix\url{https://doi.org/10.1145/1961189.1961199}.

\bibitem{randomforest}
\bibinfo{author}{Breiman, L.}
\newblock \bibinfo{title}{Random forests}.
\newblock \emph{\bibinfo{journal}{Machine Learning}}
  \textbf{\bibinfo{volume}{45}}, \bibinfo{pages}{5--32} (\bibinfo{year}{2001}).
\newblock \urlprefix\url{https://doi.org/10.1023/A:1010933404324}.

\bibitem{xgboost}
\bibinfo{author}{{Chen}, T.} \& \bibinfo{author}{{Guestrin}, C.}
\newblock \bibinfo{title}{{XGBoost: A Scalable Tree Boosting System}}.
\newblock \emph{\bibinfo{journal}{arXiv e-prints}}
  \bibinfo{pages}{arXiv:1603.02754} (\bibinfo{year}{2016}).
\newblock \eprint{1603.02754}.

\bibitem{asnn}
\bibinfo{author}{Tetko, I.~V.}
\newblock \bibinfo{title}{Associative neural network}.
\newblock \emph{\bibinfo{journal}{Neural Processing Letters}}
  \textbf{\bibinfo{volume}{16}}, \bibinfo{pages}{187--199}
  (\bibinfo{year}{2002}).
\newblock \urlprefix\url{https://doi.org/10.1023/A:1019903710291}.

\bibitem{sosnin}
\bibinfo{author}{Sosnin, S.}, \bibinfo{author}{Karlov, D.},
  \bibinfo{author}{Tetko, I.~V.} \& \bibinfo{author}{Fedorov, M.~V.}
\newblock \bibinfo{title}{Comparative study of multitask toxicity modeling on a
  broad chemical space}.
\newblock \emph{\bibinfo{journal}{Journal of Chemical Information and
  Modeling}} \textbf{\bibinfo{volume}{59}}, \bibinfo{pages}{1062--1072}
  (\bibinfo{year}{2019}).
\newblock \urlprefix\url{https://doi.org/10.1021/acs.jcim.8b00685}.
\newblock \eprint{https://doi.org/10.1021/acs.jcim.8b00685}.

\bibitem{arras}
\bibinfo{author}{Arras, L.}, \bibinfo{author}{Montavon, G.},
  \bibinfo{author}{M{\"u}ller, K.-R.} \& \bibinfo{author}{Samek, W.}
\newblock \bibinfo{title}{Explaining recurrent neural network predictions in
  sentiment analysis}.
\newblock In \emph{\bibinfo{booktitle}{Proceedings of the 8th Workshop on
  Computational Approaches to Subjectivity, Sentiment and Social Media
  Analysis}}, \bibinfo{pages}{159--168} (\bibinfo{publisher}{Association for
  Computational Linguistics}, \bibinfo{address}{Copenhagen, Denmark},
  \bibinfo{year}{2017}).
\newblock \urlprefix\url{https://www.aclweb.org/anthology/W17-5221}.

\bibitem{mutagen}
\bibinfo{author}{Plo{\v s}nik, A.}, \bibinfo{author}{Vra{\v c}ko, M.} \&
  \bibinfo{author}{Dolenc, M.~S.}
\newblock \bibinfo{title}{Mutagenic and carcinogenic structural alerts and
  their mechanisms of action}.
\newblock \emph{\bibinfo{journal}{Archives of Industrial Hygiene and
  Toxicology}} \textbf{\bibinfo{volume}{67}}, \bibinfo{pages}{169--182}
  (\bibinfo{year}{2016}).
\newblock
  \urlprefix\url{https://content.sciendo.com/view/journals/aiht/67/3/article-p169.xml}.

\bibitem{solub}
\bibinfo{author}{Huuskonen, J.}
\newblock \bibinfo{title}{Estimation of aqueous solubility for a diverse set of
  organic compounds based on molecular topology}.
\newblock \emph{\bibinfo{journal}{Journal of Chemical Information and Computer
  Sciences}} \textbf{\bibinfo{volume}{40}}, \bibinfo{pages}{773--777}
  (\bibinfo{year}{2000}).
\newblock \urlprefix\url{https://doi.org/10.1021/ci9901338}.
\newblock \bibinfo{note}{PMID: 10850781},
  \eprint{https://doi.org/10.1021/ci9901338}.

\end{thebibliography}

\end{document}